\documentclass{ws-ijmpc}
\renewcommand{\wsdraftnote}{}
\usepackage[super]{cite}
\usepackage{xcolor}
\usepackage{subcaption}
\usepackage{booktabs}
\usepackage[verbose,hypertexnames=false]{hyperref}
\hypersetup{colorlinks=false,allbordercolors=blue,pdfborderstyle={/S/U/W 1}}

\begin{document}

\markboth{M. I. Khan, S. Succi \& G. Falcucci}
{Validating the Boltzmann approach to the large-eddy simulations}

\catchline{}{}{}{}{}
%%%%%%%%%%%%%%%%%%%%%%%%%%%%%%%%%%%%%%%%%%%%%%%%%%%%%%%%%%%%%%%%%%%%

\title{Validating the Boltzmann approach to the Large-Eddy simulations of forced
    homogeneous incompressible turbulence}

\author{Muhammad Idrees Khan\footnote{
        Corresponding author.}}

\address{Department of Enterprise Engineering ``Mario Lucertini'', University of Rome ``Tor Vergata'', Via del Politecnico 1, 00133 Rome, Italy\\
    muhammadidrees.khan@students.uniroma2.eu}

\author{Sauro Succi}

\address{Italian Institute of Technology, Piazzale Aldo Moro 1, 00185 Rome, Italy\\
    Department of Physics, Harvard University, 33 Oxford Street, Cambridge, Massachusetts 02138, USA}

\author{Giacomo Falcucci}

\address{Department of Enterprise Engineering ``Mario Lucertini'', University of Rome ``Tor Vergata'', Via del Politecnico 1, 00133 Rome, Italy\\
    Department of Physics, Harvard University, 33 Oxford Street, Cambridge, Massachusetts 02138, USA}

\maketitle

\begin{abstract}

    The simulation of turbulent flows remains a central challenge, as even our most powerful computers cannot
    resolve the finest scales of motion in many flows of practical interest.
    As a result, the effects of unresolved scales on large eddies must be modelled via closures and
    coarse-graining procedures. Large-eddy simulation (LES) traditionally coarse-grains Navier-Stokes equations
    using Smagorinsky's effective viscosity model. This has the merit of simplicity but fails to account for
    strong non-equilibrium effects, as they typically arise in most flows in the vicinity of solid walls, the reason
    being  that the notion of eddy viscosity assumes scale separation between small and large eddies,
    an assumption that fails for high-Reynolds flows far from equilibrium.
    The lattice Boltzmann method (LBM) offers an alternative by coarse-graining at the kinetic level, potentially
    capturing non-equilibrium effects beyond reach of hydrodynamic closures.
    This paper addresses the question as to whether LBM-Smagorinsky LES of
    forced homogeneous isotropic turbulence (FHIT) exhibits kinetic behaviour.
    We test whether the turbulent Knudsen number $K_t$, measuring scale separation, reaches
    order one (kinetic regime) or remains asymptotically small (hydrodynamic regime).
    Using reference DNS and LES on lattices of size $800^3$ and $100^3$, in lattice units (lu), at
    $Re = 2 \times 10^4$, we quantify $K_t$ via spatial maps, temporal statistics, energy spectra, and higher-order moments.
    Results show $K_t \sim O(10^{-3})$, strictly positive without negative excursions, with spectra and flatness
    following canonical LES behaviour. We conclude that despite its kinetic formulation, LBM-Smagorinsky
    LES operates in the hydrodynamic regime, with small FHIT eddies remaining in local equilibrium
    with larger ones, validating Smagorinsky viscosity and confirming that LBM-LES functions as conventional
    hydrodynamic LES while preserving lattice Boltzmann efficiency and locality.

\end{abstract}

\keywords{Lattice Boltzmann method; Large eddy simulation; Forced homogeneous isotropic turbulence; Multiple relaxation time; Smagorinsky model}

\section{Introduction}
The simulation of turbulent flows remains one of the most compelling challenges in science and engineering. Despite tremendous progress in computational methods, we are far from being able to directly simulate many high-Reynolds flows of practical relevance, such as the full aerodynamics of an aircraft. As a result, fine scales must be modelled via turbulence closures. Conventional large-eddy simulation (LES) models achieve this goal by coarse-graining the Navier-Stokes equations, with the Smagorinsky model representing unresolved eddies through an effective viscosity.
While simple and robust, such closures cannot account for strong non-equilibrium effects, as they typically
arise near boundaries.

Homogeneous isotropic turbulence (HIT) provides a cornerstone for theoretical, numerical, and experimental studies in fluid dynamics, where turbulence is assumed uniform in all directions without preferential orientation. This idealised framework, grounded in the seminal work by von Kármán and Howarth \cite{de1938statistical}, has been pivotal in advancing statistical turbulence theory. For comprehensive overviews of HIT and statistical properties at small scales, see \cite{benzi2015homogeneous,frisch1995turbulence}.

Direct Numerical Simulation (DNS) provides the most accurate method for resolving all scales of turbulence \cite{ishihara2009study}, but computational requirements scale dramatically with Reynolds number (Re) \cite{hami2021turbulence}. LES reduces computational cost relative to DNS, yet remains expensive at high Reynolds numbers and in complex geometries \cite{pope2001turbulent,meneveau2000scale}. The Lattice Boltzmann Method (LBM) offers significant computational advantages through local information storage and updates within each lattice cell \cite{succi2018lattice,kruger2017lattice} and has been successfully applied to flows in complex porous geometries \cite{falcucci2021extreme,falcucci2024adapting}.

An alternative proposed in recent decades is to perform coarse-graining directly at the mesoscale level of the Boltzmann kinetic distribution. This avoids the need of strict scale separation between resolved and unresolved eddies and, in principle, allows non-equilibrium effects to be captured beyond the reach of effective viscosity closures. Although adopted in advanced commercial solvers, systematic benchmarks of this kinetic LES approach remain rather scanty.

LBM is a numerical approach for fluid dynamics based on Boltzmann's kinetic theory, rather than the continuum assumption of Navier-Stokes \cite{kruger2017lattice,succi2018lattice}. Compared to traditional turbulence models, LBM combined with LES offers an efficient alternative, effectively resolving large turbulent structures while modeling smaller scales \cite{kang2013effect,ansumali2004kinetic,chen2003extended, latt2006lattice,latt2006numerical}.

Forced homogeneous isotropic turbulence (FHIT) serves as a key benchmark for examining statistical turbulence theories \cite{li2024stochastic,li2024synthetic}. Previous LBM studies of FHIT have explored fundamental fluid dynamics issues \cite{gkoudesnes2019evaluating,kareem2009lattice,ten2006application,chen2004expanded}, reporting energy spectra adhering to the $k^{-5/3}$ Kolmogorov scaling law \cite{kolmogorov1991local} in the inertial sub-range, along with appropriate ranges for dimensionless turbulence parameters and other important statistical features.

The objective of this work is to \emph{quantitatively determine} whether LBM-Smagorinsky LES operates in a kinetic regime $\big(\tau > \mathcal{O}(1)\big)$ or in the hydrodynamic regime $\big(\tau = \tfrac{1}{2} + \theta\big)$, where $\tau$ is the relaxation time in the LBM collision step (for LES this is the effective $\tau_e$, Sec.~\ref{subsec:LES_MRT}; for DNS $\tau = \tau_0$) and $\theta \equiv \tau - \tfrac{1}{2} \ll 1$ in the hydrodynamic case \cite{succi2020towards,toschi2005lattice}. Equivalently, the question is whether the turbulent Knudsen number $K_t$ (Sec.~\ref{sec:hydrolimit}) reaches order unity, signaling kinetic behavior, or remains small ($K_t \sim 10^{-3}$), indicating a strictly hydrodynamic regime. We address this by (i) constructing spatial maps and time-series of the effective relaxation-time offset $\tau_e - \tfrac{1}{2}$, which is the quantity that enters the hydrodynamic criterion (Sec.~\ref{sec:hydrolimit}), under iso-Reynolds DNS/LES pairing, and (ii) cross-validating with energy spectra and higher-order statistics. Our principal result is that the method \emph{consistently} remains hydrodynamic: $\tau_e - \tfrac{1}{2} = \mathcal{O}(10^{-3})$ everywhere with no negative excursions, implying stability and strict locality of the MRT collision-streaming operator; concurrently, spectra and intermittency metrics match the expected LES phenomenology without non-physical energy accumulation. Our results confirm that FHIT eddies behave analogously to molecules in a liquid: they remain in local equilibrium with larger eddies and are adequately represented by Smagorinsky eddy viscosity. This provides the first quantitative confirmation that LBM-LES of FHIT consistently operates in the hydrodynamic regime, establishing with direct evidence that LBM-Smagorinsky LES behaves as a standard hydrodynamic LES while retaining the computational and implementation advantages of the lattice Boltzmann framework.

This paper is organized as follows. Sec.~\ref{sec:methodology} introduces the numerical method, describing the LBM code and the implementation of the Smagorinsky LES model. In Sec.~\ref{sec:validation}, we validate the DNS against theoretical predictions and present a parametric analysis of the LES setup to establish accuracy and reliability. Sec.~\ref{sec:results} presents the main results: first, a detailed investigation of the effective relaxation time $\tau_e$, then a comparison between high-resolution DNS and lower-resolution LES at matched Reynolds number, and finally analysis of higher-order statistics, which collectively constitute the core contribution of this work. %Conclusions are drawn in Sec.~5.

\section{Numerical Methodology}
\label{sec:methodology}

\subsection{Lattice Boltzmann Method}
LBM is grounded on a Boltzmann-type kinetic equation over a discrete set of velocities corresponding to a space-filling lattice. It can be mathematically derived from the Boltzmann equation and recovers the Navier-Stokes (NS) equations under small departures from local equilibria and low Knudsen number $Kn$ numbers \cite{ubertini2005recent,qi2005many,chen1998lattice,benzi1992lattice}.

In this work we use the D3Q19 velocity stencil (see Fig.~\ref{fig:d3q19_table} and Fig.~\ref{fig:d3q19_stencil} in \ref{app:mrt_parameters}) and the Multiple-Relaxation-Time (MRT) formulation \cite{d2002multiple}. External forcing is incorporated with the Guo scheme \cite{guo2002discrete}. The discrete MRT evolution equation, equilibrium distribution, forcing term, and macroscopic relations are reported in \ref{app:lbe_mrt_forcing}; the orthogonal moment matrix $M$, the diagonal relaxation matrix $S$ (Eq.~\ref{eq:relaxation-matrix}), and the equilibrium/force moments (Eq.~\ref{eq:meq_fm}) are in \ref{app:mrt_parameters}.

In the MRT framework, the kinematic viscosity $\nu_0$ is controlled by the shear relaxation rate $s_\nu$ \cite{yu2006turbulent}:
\begin{equation}
    \nu_0 = c_s^2 \left( \frac{1}{s_\nu} - \frac{1}{2} \right) \delta t,
    \label{eq:bare_viscosity}
\end{equation}
with lattice speed $c=\delta x/\delta t$ and sound speed $c_s=c/\sqrt{3}$ for D3Q19 \cite{mohamad2011lattice}. This generalizes the Bhatnagar-Gross-Krook (BGK) relation and enables improved stability at high Reynolds numbers. The Guo-MRT formulation constitutes the backbone of our solver for incompressible flow.

\subsection{Forced Homogeneous Isotropic Turbulence}
\label{subsec:FHIT}

FHIT is characterized by its energy spectrum $E(k)$ and associated turbulence scales. We adopt the spectrum and integral scale definitions from \cite{kareem2009lattice}:
\begin{equation}
    E(k) = \frac{1}{2} \sum_{k - \frac{1}{2} \leq |k| < k + \frac{1}{2}} |\hat{u}(k)|, \quad
    L = \frac{3\pi}{4} \frac{\int \frac{E(k)}{k} \, dk}{\int E(k) \, dk}. \label{eq:integral_length_scale}
\end{equation}
Here, $\hat{u}(k)$ is the Fourier transform of the velocity field $u(x)$, and $L$ represents the characteristic large-eddy size, which can be interpreted as the energy-weighted average of the inverse wavenumber, $L \sim \langle 1/k \rangle$.
Kolmogorov scale $\eta$, Taylor microscale $\lambda$, root mean square velocity $u_{\mathrm{rms}}$, and dissipation rate $\varepsilon$ are given by \cite{yu2005dns}:
\begin{align}
    \eta             & = \left( \frac{\nu^3}{\varepsilon} \right)^{1/4}, \label{eq:kolmogorov_scale}        \\
    \lambda          & = \sqrt{\frac{15 \nu u_{\mathrm{rms}}^2}{\varepsilon}}, \label{eq:taylor_microscale} \\
    u_{\mathrm{rms}} & = \sqrt{\frac{2k}{3}}, \label{eq:rms_velocity}                                       \\
    \varepsilon      & = 2\nu \int k^2 E(k) \, dk. \label{eq:dissipation_rate}
\end{align}

To sustain FHIT in a periodic cube, we apply a stationary, divergence-free, spatially periodic forcing $F = (F_x,F_y,F_z)$ with components defined in Eq.~\ref{eq: forcing_pattern}.
In the numerical implementation, this forcing field is used as the external force $F^{ext}$ within the Guo-MRT formulation (\ref{app:lbe_mrt_forcing}):

\begin{equation}
    \begin{aligned}
        F_x(i,j,k) & = F_0 \sin\!\left( \frac{2\pi j}{N_y} \right), \\
        F_y(i,j,k) & = F_0 \sin\!\left( \frac{2\pi k}{N_z} \right), \\
        F_z(i,j,k) & = F_0 \sin\!\left( \frac{2\pi i}{N_x} \right),
    \end{aligned}
    \label{eq: forcing_pattern}
\end{equation}
with grid resolution $N_x=N_y=N_z$ and
\begin{equation}
    F_0 = C_f \cdot \frac{\rho U_0^2}{N_x},
    \label{eq:forcing_amplitude}
\end{equation}
where $C_f=0.2$, $\rho$ is the fluid density, and $U_0$ the characteristic velocity \cite{zhang2024subgrid}. Initial conditions are Taylor-Green vortices (TGV) \cite{brachet1983small}. Energy balance is verified in real space via
\begin{equation}
    \varepsilon = -\frac{dE_k}{dt} + \langle F \cdot u \rangle.
    \label{eq:energy_balance}
\end{equation}
where $E_k = \frac{1}{2} \langle u_i u_i \rangle$ is the turbulent kinetic energy, $S_{ij} = \frac{1}{2}(\partial u_i/\partial x_j + \partial u_j/\partial x_i)$ is the strain rate tensor, and $\varepsilon = 2\nu_0 \langle S_{ij} S_{ij} \rangle$ is the dissipation rate.

%%%%%%%%%%%%%%%%%MRT LES %%%%%%%%%%%%%%%%%%%%%%%%%%%%%%%%
\subsection{LES-Based Extension of LBM-MRT}
\label{subsec:LES_MRT}

Extending the LBM to an LES framework allows coarser lattice resolutions by modeling unresolved scales, so that high-Reynolds turbulence can be simulated when full DNS is prohibitively expensive; on such coarser grids, the subgrid model also improves numerical stability and physical accuracy relative to under-resolved DNS without a model \cite{hou1994lattice,guo2018applications,yu2005dns,dong2008inertial,chen2009large,an2020new}. In this work, we achieve this extension by incorporating the Smagorinsky eddy-viscosity model into MRT-LBM, following \cite{yu2006turbulent}. All fields are implicitly filtered in LBM-LES. The total effective viscosity and the Smagorinsky closure are

\begin{equation}
    \nu_e = \nu_0 + \nu_t,
    \label{eq:nu_eff_mrt}
\end{equation}
\begin{equation}
    \nu_t = C \Delta^2 |S|,
    \label{eq:eddy_visc}
\end{equation}
where $\Delta$ is the LES filter width (in LBM-LES taken as the grid spacing $\delta x$), $C$ is the model constant, and $|S|$ is the Frobenius norm of the local strain-rate tensor. The corresponding effective relaxation time is
\begin{equation}
    \frac{1}{s_\nu} = \frac{1}{2} + 3(\nu_0+\nu_t) \equiv \tau_e.
    \label{eq:les_viscosity}
\end{equation}
Only the shear-related MRT entries are updated locally based on $|S|$; other diagonal entries remain fixed \cite{yu2006turbulent}. The six strain-rate components $S_{ab}$ are computed from non-equilibrium moments as in \ref{app:strain_rate_tensor} (Eq.~\ref{eq:yu_strain}); the norm is
\begin{equation}
    |S| = \sqrt{2 \left( S_{xx}^2 + S_{yy}^2 + S_{zz}^2 + 2S_{xy}^2 + 2S_{xz}^2 + 2S_{yz}^2 \right)}.
    \label{eq:strain_norm}
\end{equation}

The LES-MRT evolution equation with forcing, consistent with the DNS-MRT form \ref{eq:mrt_dns_guo}, is
\begin{align}
    f(x + c_\alpha \delta t, t + \delta t) - f(x, t) = {} &
    - \left[ M^{-1} \Lambda(\nu_e) M (f - f^{eq}) \right] \nonumber                                                                                \\
                                                          & + \delta t \left[ M^{-1} \left( I - \frac{1}{2} \Lambda(\nu_e) \right) M \right] \Phi,
    \label{eq:mrt_les_guo}
\end{align}
where $M$ is the moment transformation matrix, $\Lambda(\nu_e)$ is the diagonal relaxation matrix with effective viscosity $\nu_e$, and $\Phi$ is the discrete source term (see \ref{app:lbe_mrt_forcing} for complete definitions).
This framework reduces to DNS-MRT as $C \to 0$ (cf. Fig.~\ref{fig:c_variation}) and preserves LBM locality. The hydrodynamic limit requires the relaxation-time offset $\tau - \tfrac{1}{2}$ to be small (Sec.~\ref{sec:hydrolimit}); we therefore use $\tau_e - \tfrac{1}{2}$ as the diagnostic to identify the operating regime.
This diagnostic can be expressed in terms of viscosity and grid parameters as
\begin{equation}
    \tau_e - \tfrac{1}{2} \;=\; \frac{\nu_e}{c_s^2}
    \;=\; \frac{\nu_0+\nu_t}{1/3}
    \;=\; 3\!\left(\frac{U_0 N_x}{Re} + C\,\Delta^2 |S|\right),
    \label{eq:tau_offset_scaling}
\end{equation}
which highlights the contributions of the base viscosity $\nu_0$ and the Smagorinsky eddy viscosity $\nu_t$ to the offset from the hydrodynamic regime, defined by $(\tau - \tfrac{1}{2}) \ll 1$.
As discussed in Sec.~\ref{sec:hydrolimit}, hydrodynamic consistency requires
$(\tau - \tfrac{1}{2}) \ll \ell/(\sqrt{3}\,\delta x)$, where $\ell$ is the smallest characteristic length scale of the resolved field.
In the LES extension, the same criterion applies with the effective relaxation time $\tau_e$,
so the diagnostic $\tau_e - \tfrac{1}{2}$ directly verifies whether the simulation remains in the hydrodynamic regime.

\subsection{Hydrodynamic Limit}
\label{sec:hydrolimit}

The lattice Boltzmann method recovers the Navier-Stokes equations in the hydrodynamic limit, which requires that the Knudsen number
\begin{equation}
    Kn \equiv \frac{\lambda}{\ell} \ll 1
\end{equation}
remains small \cite{ansumali2004kinetic,succi2001lattice}. Here, $\lambda = (\tau - \tfrac{1}{2}) \sqrt{3}\,\delta x$ is the lattice mean free path and $\delta x$ is the grid spacing. The length $\ell$ is the smallest characteristic scale that the scheme resolves: in DNS this is the physical smallest flow scale (e.g.\ Kolmogorov length $\eta$ in turbulence, or viscous length in wall-bounded flows); in LES it is the smallest scale in the resolved field, i.e.\ the filter scale (in this work $\Delta = \delta x$).

This condition can be written equivalently as
\begin{equation}
    \tau - \tfrac{1}{2} \ll \frac{\ell}{\sqrt{3}\,\delta x}.
\end{equation}
If $\tau - 1/2 = \mathcal{O}(1)$, the scheme enters the kinetic regime and hydrodynamic consistency is lost. In contrast, offsets of order $10^{-3}$-$10^{-2}$ ensure that the mean free path remains vanishingly small compared to $\ell$, preserving the hydrodynamic regime. For LES, the effective relaxation time $\tau_e$ (Sec.~\ref{subsec:LES_MRT}) replaces $\tau$; the corresponding \emph{turbulent Knudsen number} is
\begin{equation}
    K_t \equiv \frac{\lambda_e}{\ell}, \qquad \lambda_e = (\tau_e - \tfrac{1}{2}) \sqrt{3}\,\delta x,
    \label{eq:Kt_definition}
\end{equation}
so that $K_t \ll 1$ corresponds to the hydrodynamic regime and $K_t \sim \mathcal{O}(1)$ to the kinetic regime.

For HIT specifically: in DNS, $\ell \sim \eta$ with $\eta \gtrsim \delta x$ when the grid resolves the dissipative range; in LES, $\ell \sim \delta x$ by the above definition. In both cases $\ell \gtrsim \delta x$, so that
\begin{equation}
    \tau - \tfrac{1}{2} \ll \mathcal{O}(1).
    \label{eq:hydro_condition}
\end{equation}
Thus non-hydrodynamic effects are not expected in FHIT, whereas in wall-bounded flows with strong shear near the walls such effects may become relevant.

%%%%%%%%%%%%%%%%%%%%%%%%%%%%%
\section{Validation}
\label{sec:validation}

\subsection{Simulation Setup}

An LBM-based Computational Fluid Dynamics (CFD) solver was developed in Fortran. This solver is implemented in serial (single core), and OpenACC (GPU parallelism), with all simulations in the present research work conducted using the OpenACC version. The macroscopic quantities are recorded at specific iterations for subsequent post-processing. Post-processing data is computed on the fly at set intervals. Energy spectra are computed using GPU-accelerated  Fast Fourier Transform  (cuFFT) library \cite{nvidia_cufft}.

All quantities in this work are given in lattice units (lu). We set $t_0 = 1000$ so that $t/t_0$ is the time in units of 1000 iterations. The computational domain for the FHIT simulation is a $2\pi$ periodic box in all three spatial directions, with periodic boundary conditions applied throughout. The initial velocity field is initialized using the $TGV$ configuration~\cite{brachet1983small}, consistent with prior FHIT simulations. Simulations are performed and compared with reference Reynolds number $Re = \frac{U_0 N_x}{\nu_0}$ where $U_0=0.043$ \cite{zhang2024subgrid} and the Taylor-scale Reynolds number $Re_\lambda = \frac{u_{rms} \lambda}{\nu_0}$ \cite{ishihara2009study}.

The fluid is assumed to be incompressible ($\nabla \cdot u=0$) and we perform both DNS and LES of FHIT using different grid sizes ($N_{\text{x, DNS}}^3$ and $N_{\text{x, LES}}^3$). For the LES, the grid size is reduced by a constant factor in each direction compared to the DNS grid size (by a factor of 8 in this study, Sec.~\ref{sec:results}, e.g., a DNS grid of $800^3$ corresponds to an LES grid of $100^3$). This approach allows for results that are comparable to those from DNS at higher resolutions, while significantly reducing computational time and memory usage.

\begin{figure*}[b]
    \centerline{\includegraphics[width=9.4cm]{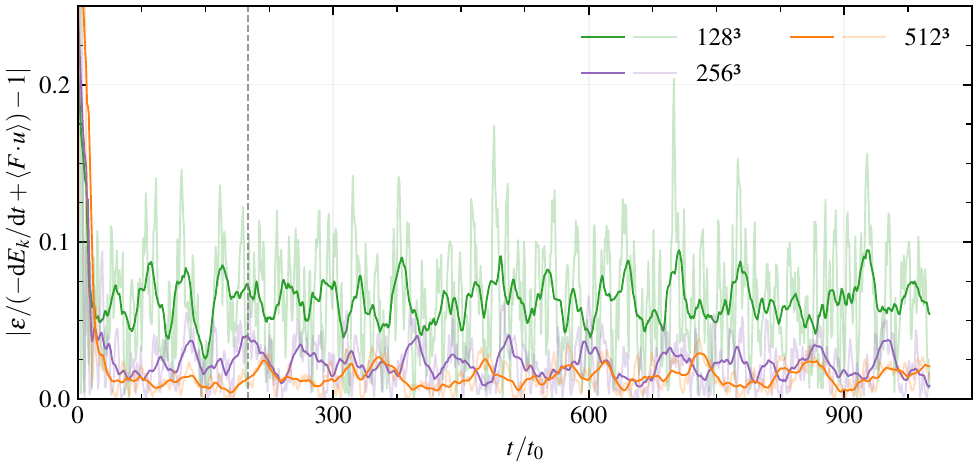}}
    \vspace*{8pt}
    \caption{Temporal evolution of the relative error for different grid resolutions: $128^3$ (green), $256^3$ (purple), $512^3$ (orange). Faint lines show raw instantaneous values, while solid lines represent the 20-point rolling average. The vertical dashed line at $t/t_0 = 200$ marks the start of the statistically stationary period used for turbulence post-processing.}
    \label{fig:energy_balance}
\end{figure*}

Consistency of the energy injection mechanism is verified through the energy balance error discussed in Sec.~\ref{subsec:FHIT} Eq.~\ref{eq:energy_balance}. For simulations at $Re = 2500$ for $128^3$, $256^3$ grids, and $Re = 5000$ for $512^3$ grid the relative error stabilizes after $t/t_0 = 200$ at approximately $6-7\%$ for the $128^3$ grid, below $3.5\%$ for $256^3$, and around $1\%$ for the $512^3$ resolution. These results confirm the accuracy of the forcing scheme and its consistent coupling with the energy evolution. All turbulence statistics and post-processing results presented in this study are computed using data from the statistically stationary interval $t/t_0 \in [200, 1000]$, as marked by the vertical dashed line in Fig.~\ref{fig:energy_balance}.

\subsection{Assessment of Statistical Isotropy}
\label{subsec:isotropy}

\begin{figure}[b]
    \centerline{\includegraphics[width=9.4cm]{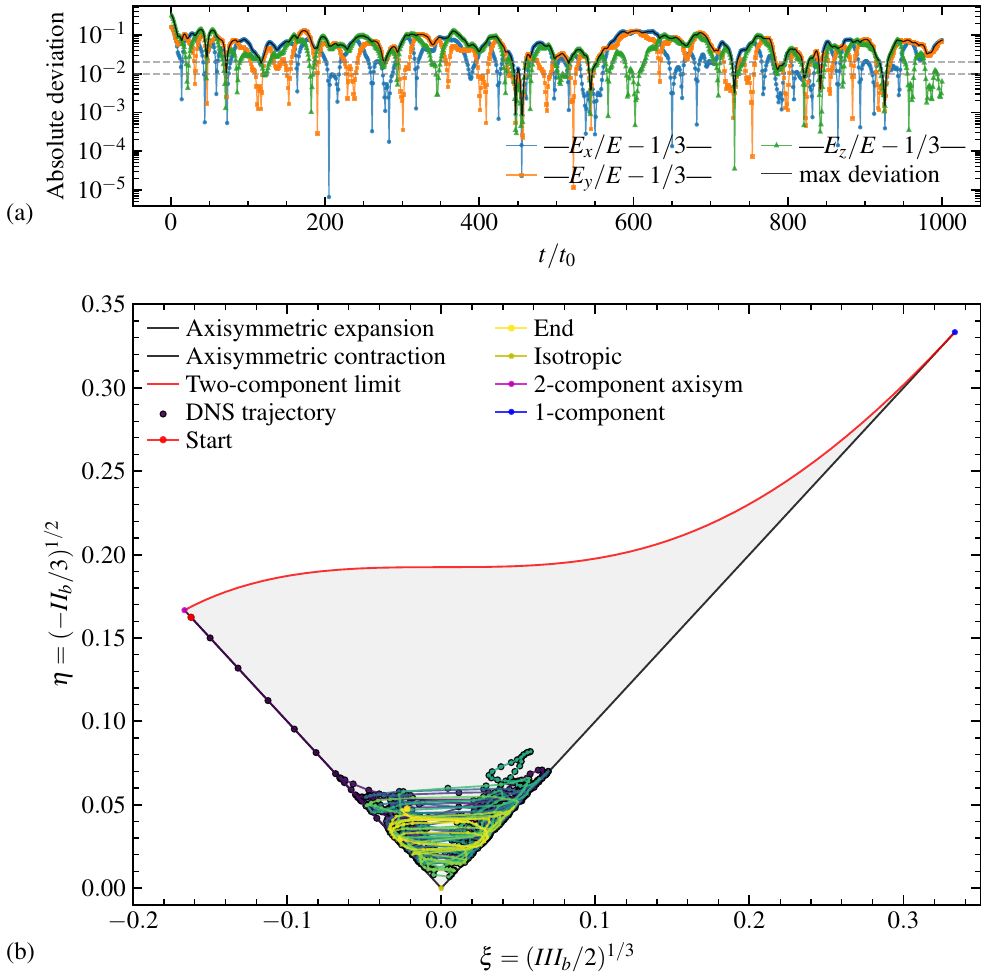}}
    \vspace*{8pt}
    \caption{Assessment of statistical isotropy.
        (a) Time history of the absolute deviation of directional energy fractions from the isotropic value $1/3$. The peak at the left corresponds to the initial condition ($t=0$).
        (b) Lumley triangle: trajectory of the Reynolds stress anisotropy invariants. Coordinates are $\xi = (III_b/2)^{1/3}$ and $\eta = (-II_b/3)^{1/2}$, where $II_b$ and $III_b$ are the second and third invariants of the Reynolds stress anisotropy tensor $b_{ij}$ \cite{pope2001turbulent}. The legend distinguishes between simulation data and the theoretical background map: (1) \textbf{Theoretical boundaries \& vertices:} The solid lines (``Axisymmetric expansion'', ``Axisymmetric contraction'', ``Two-component limit'') and vertices (``1-component'', ``2-component axisym'', ``Isotropic'') denote the limits of realizable turbulence \cite{yang2021return,pope2001turbulent}. (2) \textbf{Simulation data:} The ``DNS trajectory'' (dots) tracks the evolution from ``Start'' (red marker) to the final state (``End''), which converges to the isotropic vertex. Statistical isotropy is validated for the $512^3$ DNS at ${Re}_\lambda \approx 180$, $\eta^+ \approx 1.002$.}
    \label{fig:isotropy_validation}
    \alttext{Assessment of statistical isotropy showing absolute deviations of directional energy fractions and the Lumley anisotropy invariant map.}
\end{figure}

\begin{table}[htbp]
    \tbl{Summary of isotropy validation metrics from DNS FHIT.
        Energy fractions are normalized by the total kinetic energy.\label{tab:isotropy_metrics}}
    {\begin{tabular}{@{}lcc@{}}
            \toprule
            \textbf{Metric}       & \textbf{Value} & \\
            \colrule
            Ex/E mean             & 0.3431         & \\
            Ex/E std              & 0.0490         & \\
            Ey/E mean             & 0.3316         & \\
            Ey/E std              & 0.0538         & \\
            Ez/E mean             & 0.3253         & \\
            Ez/E std              & 0.0563         & \\
            max $|E_i/E - 1/3|$   & 0.3248         & \\
            mean $|b_{12}|$       & 0.0161         & \\
            mean $|b_{13}|$       & 0.0157         & \\
            mean $|b_{23}|$       & 0.0156         & \\
            mean anisotropy index & 0.0978         & \\
            \botrule
        \end{tabular}}
\end{table}

The statistical isotropy of the simulated turbulent field is rigorously evaluated using the Reynolds stress anisotropy tensor $b_{ij}$ and complementary diagnostics (foundations in \cite{pope2001turbulent}). As shown in Fig.~\ref{fig:isotropy_validation}(a), the absolute deviations $|E_i/E_{\text{tot}} - 1/3|$ (where $E_i = E_x, E_y, E_z$) and their maximum (black curve) remain bounded within a $\pm 1\text{-}2\%$ tolerance after the initial transient (horizontal dashed lines at 1\% and 2\%), with the largest excursions occurring in the $z$-component. Fig.~\ref{fig:isotropy_validation}(b) shows the Lumley triangle, which maps the invariants $II_b$ and $III_b$ of $b_{ij}$ onto coordinates $\xi$ and $\eta$ \cite{pope2001turbulent}. The top curve is the two-component limit (turbulence confined to a plane); the left and right boundaries represent axisymmetric contraction and axisymmetric expansion, respectively. The DNS trajectory (colored by simulation time from blue to yellow) tracks the evolution from the initial TGV state (red marker) toward the isotropic vertex at $(0,0)$ (yellow marker: final state), confirming that the forcing maintains statistical isotropy.

The quantitative summary in Table~\ref{tab:isotropy_metrics} confirms that the turbulence is statistically isotropic in the mean. The time-averaged directional energy fractions differ from $1/3$ by less than $1\%$ (e.g.\ $E_x/E \approx 0.34$), while the standard deviations remain small ($\approx 0.05$). The maximum absolute deviation of 0.325 corresponds strictly to the initial condition (seen at $t/t_0 = 0$ in Fig.~\ref{fig:isotropy_validation}a); following this transient, the deviations drop significantly and the flow relaxes to a state where off-diagonal anisotropy tensor components remain below $\mathcal{O}(10^{-2})$ and the mean anisotropy index remains below 0.1. Collectively, these diagnostics confirm that the forced DNS successfully maintains statistical isotropy. The spectral-space isotropy for the $512^3$ case is quantitatively validated through the isotropy coefficient analysis in \ref{app:spectral_isotropy}.

\subsection{Energy Cascade}
All energy spectra presented in this work—both DNS and LES—are time-averaged over the interval $t/t_0 = 200 - 1000$, during which the flow reaches a statistically steady state. Shaded regions around the spectra indicate one standard deviation from the mean, capturing temporal variability across snapshots.
\subsubsection{DNS spectral analysis}
\label{subsec:energy_cascade}
To support the validity of our simulation, we examine the energy spectra $E(k)$ obtained from the DNS. The results show a clear inertial subrange that aligns with the theoretical $k^{-5/3}$ slope predicted by Kolmogorov's theory. This agreement confirms that the energy cascade is accurately captured by our DNS, with proper development of homogeneous isotropic turbulence.

To verify the results obtained via spectral analysis of the DNS, we employ the spectral framework developed in \cite{pope2001turbulent}. This model synthesizes empirical results from numerous experimental studies, offering a unified representation of the energy spectrum $E(k)$:

\begin{equation}
    E(k) = C_K \epsilon^{2/3} k^{-5/3} f_L(k L) f_\eta(k \eta) \, ,
    \label{eq:pope_model}
\end{equation}

\begin{figure}[b]
    \centerline{\includegraphics[width=9.4cm]{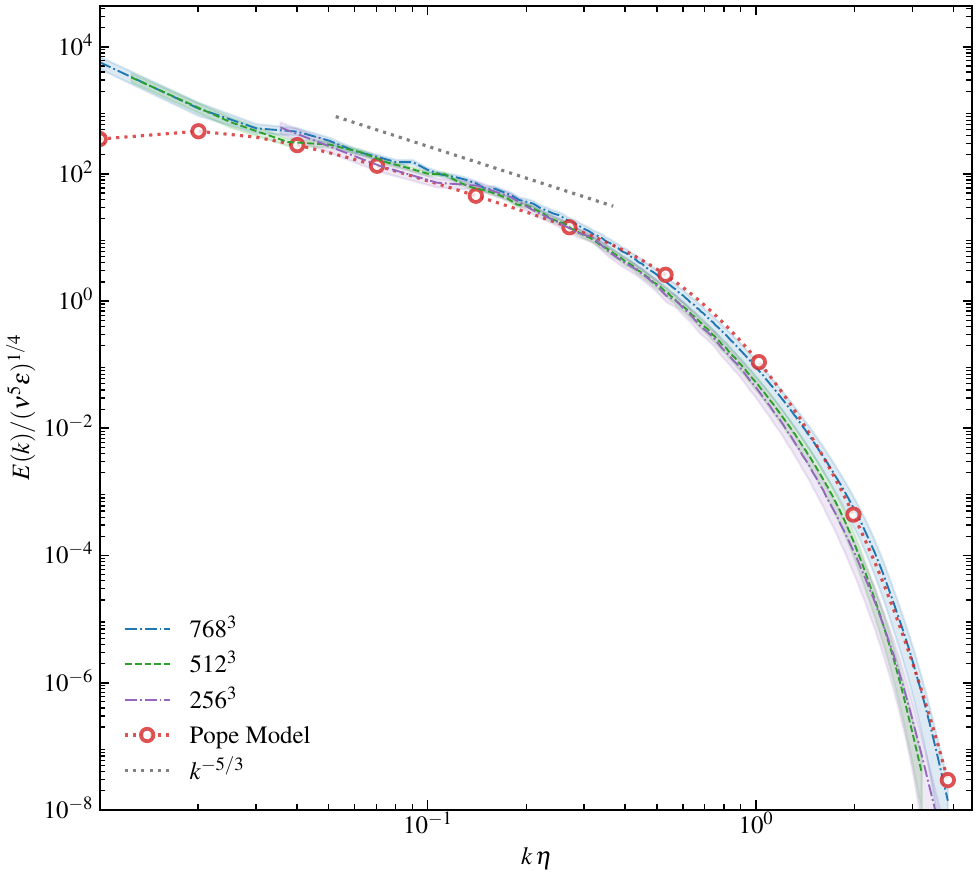}}
    \vspace*{8pt}
    \caption{Validation of DNS energy spectra against Pope model (Eq. \ref{eq:pope_model}) using normalized wavenumber $k\eta$ and Kolmogorov-scaled energy $E(k)/(\nu^5\varepsilon)^{1/4}$. Collapse across resolutions and $Re_\lambda$ confirms grid independence and $k^{-5/3}$ scaling.}
    \label{fig:DNS_spectra_validation}
    \alttext{Validation of DNS energy spectra against Pope model showing normalized wavenumber and Kolmogorov-scaled energy.}
\end{figure}

\noindent where $C_K=1.5$ is the Kolmogorov constant, $k$ denotes the wavenumber, and $\epsilon$ represents the turbulent kinetic energy dissipation rate (Eq.~\ref{eq:dissipation_rate}).
The dimensionless functions $f_L(\kappa L)$ and $f_\eta(\kappa \eta)$ modulate the spectral shape in the energy-containing (large-scale) and dissipation (small-scale) regimes, respectively. Here, $L$ corresponds to the integral length scale of dominant eddies (Eq.~\ref{eq:integral_length_scale}), while $\eta$ is the Kolmogorov microscale defining the dissipative range (Eq.~\ref{eq:kolmogorov_scale}). A detailed derivation and discussion of the model are provided in \cite{pope2001turbulent}.

Fig.~\ref{fig:DNS_spectra_validation} shows that the normalized $E(k)$ from simulations at different resolutions and $Re_\lambda$ collapsed onto the Pope model, confirming grid-independent convergence and accurate inertial-range scaling. Minor deviations at the largest scales were attributed to $Re_\lambda$ dependence and domain-forcing effects~\cite{gkoudesnes2019evaluating}. We assessed the mesh adequacy using the dimensionless Kolmogorov length scale, $\eta^+ = \frac{\eta}{\delta x}$~\cite{zhang2024subgrid}, which remained $\gtrsim 1$ across all cases \footnote{except the  $128^3$ grid in Sec.~\ref{subsec:les_spectral_analysis}, where $\frac{\eta}{\delta x}=0.43$ serves as a reference for the LES sensitivity analysis at $Re$ $5000$}, ensuring that dissipative scales were well resolved in accordance with DNS resolution standards. The specific values for all DNS simulations are summarized in Table~\ref{tab:eta_dx_values}.

\begin{table}[ht]
    \tbl{DNS resolution adequacy: dimensionless Kolmogorov length scale $\eta^+ = \eta/\delta x$ for different grid resolutions.\label{tab:eta_dx_values}}
    {\begin{tabular}{@{}cccc@{}}
            \toprule
            Grid size & $\eta/\delta x$ & $Re_\lambda$ & \\
            \colrule
            $256^3$   & 1.432           & 84           & \\
            $512^3$   & 1.672           & 127          & \\
            $512^3$   & 1.002           & 180          & \\
            $768^3$   & 1.093           & 239          & \\
            $800^3$   & 1.001           & 243          & \\
            \botrule
        \end{tabular}}
\end{table}

\subsubsection{LES spectral analysis}
\label{subsec:les_spectral_analysis}

\begin{figure}[b]
    \centerline{\includegraphics[width=9.4cm]{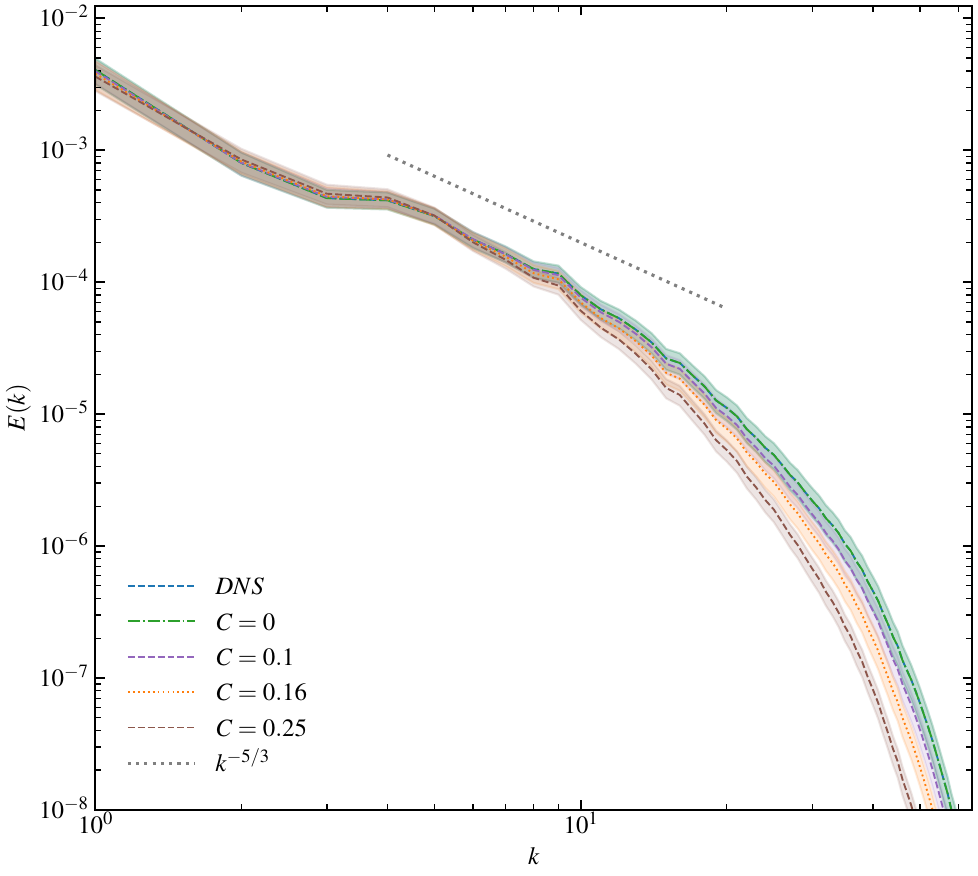}}
    \vspace*{8pt}
    \caption{Energy spectra $E(k)$ at $Re = 5000$ comparing reference DNS (blue dashed line) with LES on a $128^3$ grid using varying Smagorinsky constants $C$. With $C=0$ (green), the eddy viscosity vanishes; the LES spectrum collapses to the DNS profile at the same grid size and viscosity. Increasing $C$ to $0.1$ (purple) and $0.16$ (orange) yields intermediate dissipation levels; $C=0.25$ (brown) results in over-dissipation that suppresses small-scale energy. The theoretical Kolmogorov scaling $k^{-5/3}$ is shown for reference.}
    \label{fig:c_variation}
\end{figure}

The LES approach is validated through a parametric analysis of the model coefficient $C$. As shown in Fig.~\ref{fig:c_variation}, setting $C = 0$ reverts the solver to a DNS framework. On this coarse grid, this results in energy accumulation (pile-up) at high wavenumbers, whereas $C=0.25$ results in over-dissipation that suppresses small-scale energy. Consequently, $C = 0.16$ is identified as a balanced choice within this range.

For our primary simulation at $Re = 20000$ (Sec.~\ref{sec:spectral_analysis}), we employ $C = 0.13$, which maintains accuracy while ensuring numerical stability at higher Re. A comparison between $C = 0.16$ and the DNS reference at high Reynolds number is provided in \ref{fig:combined_dns_les_spectra} (b).

\subsection{Structure functions and intermittency metrics}

The longitudinal velocity increment is defined as
\begin{equation}
    \delta_r u = [u(x + r) - u(x)] \cdot r/|r|,
    \quad r = |r|.
    \label{eq:velocity_increment}
\end{equation}
We compute the structure functions and flatness as
\begin{align}
    S_p(r) & = \langle (\delta_r u)^p \rangle, \label{eq:structure_function} \\
    F_4(r) & = \frac{S_4(r)}{[S_2(r)]^2}, \label{eq:flatness}
\end{align}
where $\langle \cdot \rangle$ denotes spatiotemporal averaging.

\begin{figure}[b]
    \centerline{\includegraphics[width=9.4cm]{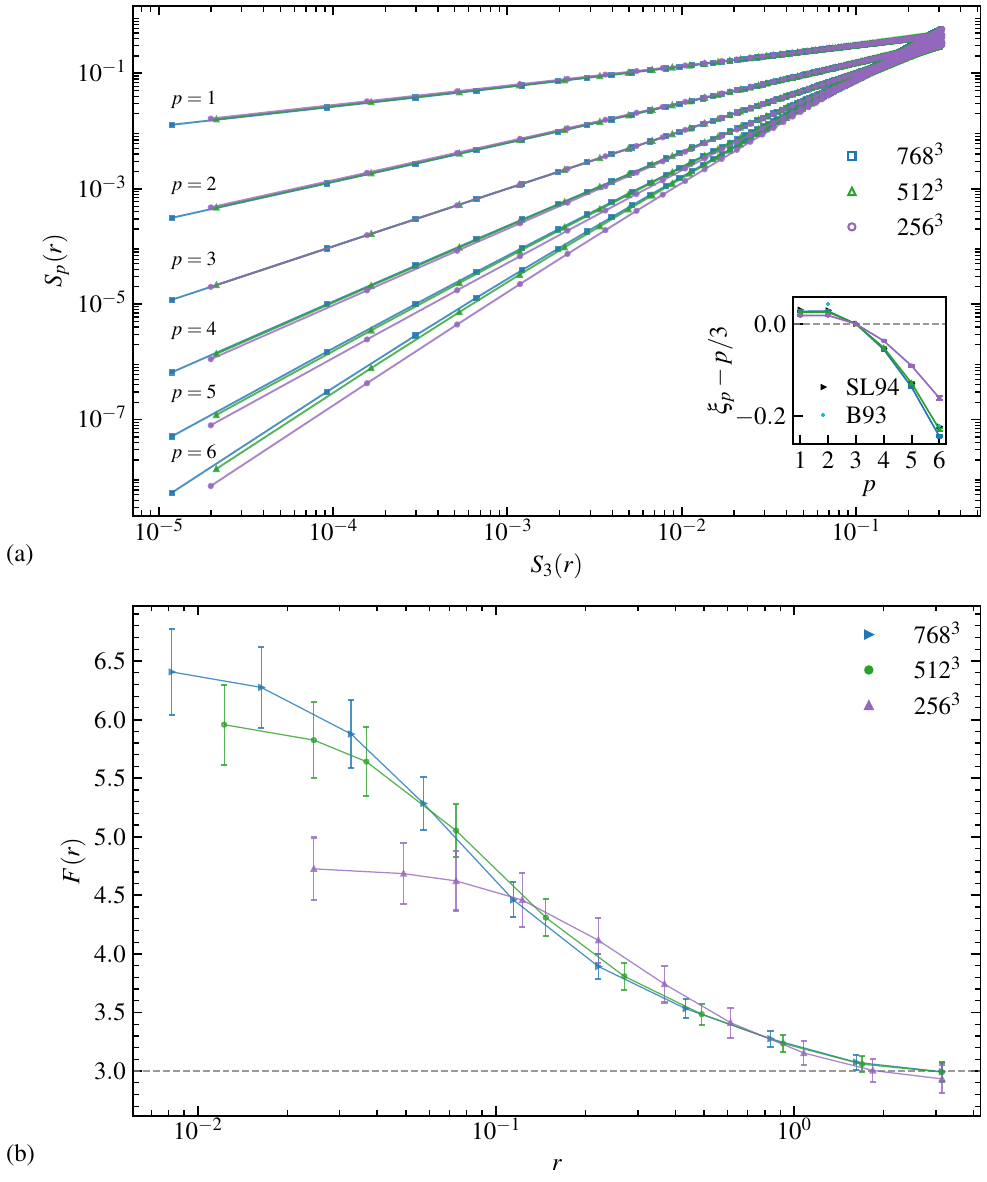}}
    \vspace*{8pt}
    \caption{Turbulence intermittency quantification.
    (a) ESS scaling $S_p \sim S_3^{\zeta_p}$ reveals anomalous exponents $\zeta_p$ (inset: comparison with She-Leveque model (SL94) and experimental anomalies of Benzi et al. (B93)).
    (b) Flatness $F_4(r)$ exceeding the Gaussian value of 3 indicates small-scale intermittency.}
    \label{fig:dns_flatness_structure}
\end{figure}

Key results are reported in Fig.~\ref{fig:dns_flatness_structure}.
Extended self-similarity (ESS) (Fig.~\ref{fig:dns_flatness_structure}) (a) provides a refined scaling analysis, revealing anomalous exponents $\zeta_p$ from $S_p \sim S_3^{\zeta_p}$, in good agreement with the She-Leveque intermittency model (SL94) \cite{she1994universal} and with experimental anomalies reported by Benzi et al. (B93) \cite{benzi1993extended}.
The flatness $F_4(r)$ (Fig.~\ref{fig:dns_flatness_structure}) (b) exhibits strong small-scale intermittency ($F_4 \gg 3$) and relaxes toward the Gaussian value ($F_4=3$) at large scales.

\section{Results and Discussion}
\label{sec:results}

The central objective of this study is to determine whether the Smagorinsky LES implemented in the MRT-LBM framework operates in the kinetic regime ($\tau>\mathcal{O}(1)$) or in the hydrodynamic regime ($\tau = \tfrac{1}{2}+\theta$ with $\theta \ll 1$). Our analysis shows that the latter holds: measured effective relaxation offsets are $\tau_e - \tfrac{1}{2} \approx 1.34\times10^{-3}$, which is nearly three orders of magnitude smaller than the hydrodynamic-limit bound given in Eq.~(\ref{eq:hydro_condition}) of Sec.~\ref{sec:hydrolimit}.

For the present LES setup with $\delta x=1$, the smallest scale in the resolved field is the grid scale, so $\ell \sim \delta x$ (Sec.~\ref{sec:hydrolimit}). Hence the hydrodynamic condition from Eq.~(\ref{eq:hydro_condition}) becomes
\[
    \tau_e - \tfrac{1}{2} \ll \mathcal{O}(1).
\]
The measured offsets are $\tau_e - \tfrac{1}{2} \approx 1.34\times10^{-3} \ll 1$, confirming that the scheme operates strictly within the hydrodynamic regime. As a result, collision-streaming remains local and turbulence statistics are reproduced faithfully.

Previous studies have implemented Smagorinsky models in both BGK- and MRT-based LBM, demonstrating that they can reproduce canonical turbulence statistics in benchmark flows such as homogeneous isotropic turbulence and turbulent jets \cite{yu2006turbulent,yu2005dns,yu2005lattice,dong2008inertial,geller2013turbulent,liu2023large,spinelli2024key}, while more recent approaches employ machine learning to optimize these closures \cite{fischer2025optimal}. However, these works focus on flow physics and model performance rather than on the operating regime of the relaxation time itself.
To the best of our knowledge, a systematic quantification of $\tau_e - \frac{1}{2}$ through spatial maps, temporal statistics, and histograms—used here to distinguish hydrodynamic ($\tau = \frac{1}{2}+\theta$) from kinetic ($\tau=\mathcal{O}(1)$) operation—has not been documented for FHIT. This forms the distinctive contribution of the present study.

To assess the physical consistency of the MRT-based LES formulation introduced in our numerical setup, we examine whether the scheme preserves the spatial locality inherent to LBM and reproduces key turbulence characteristics. Our assessment has three parts: (i) a direct, \emph{a priori} diagnosis of the operating regime through the effective relaxation time (Sec.~\ref{sec:effective_relaxation_time_analysis}), (ii) comparison of energy spectra under iso-Reynolds scaling (Sec.~\ref{sec:spectral_analysis}), and (iii) higher-order statistics (Sec.~\ref{sec:high_order_stats}).

We perform a reference DNS at $Re = 20000$, using lattice Boltzmann parameters: grid resolution $N_{x, DNS}^3 = 800^3$, $U_{0, DNS} = 0.043$, and base viscosity $\nu_{0, DNS} = 0.00172$. The corresponding LES adopts a coarser grid $N_{x, LES}^3 = 100^3$, preserving the same velocity, $U_{0, LES} = 0.043$. To maintain $Re$ similarity, $\nu_0$ is scaled as $\nu_{0,LES} = \nu_{0, DNS} \cdot (N_{x, LES} / N_{x, DNS}) = 0.000215$, yielding:
\begin{equation}
    Re_{DNS} = \frac{U_{0, DNS} N_{x, DNS}}{\nu_{0, DNS}} = 20000, \quad
    Re_{LES} = \frac{U_{0, LES} N_{x, LES}}{\nu_{0,LES}} = 20000.
\end{equation}

In the MRT formulation, the kinematic viscosity is related to the shear relaxation time via $\nu_0 = c_s^2 \left( \tau_0 - \frac{1}{2} \right)$ where $\tau_0 = 1/s_\nu$ (from Eq. \ref{eq:bare_viscosity}). This gives the LES molecular relaxation offset:
\begin{equation}
    \tau_{0,LES} - \frac{1}{2} = \frac{\nu_{0,LES}}{c_s^2} = 0.000645,
\end{equation}
which is consistent with the DNS offset $\tau_{0,DNS} - \frac{1}{2} = 0.00516$, maintaining the scaling ratio $N_{x, DNS} / N_{x, LES} = 8$.

To account for subgrid-scale effects, the Smagorinsky formulation for eddy viscosity (Eq.~\ref{eq:eddy_visc}) is employed, with model constant $C = 0.13$.
Based on inertial-range turbulence scaling arguments \cite{kolmogorov1991dissipation}, the strain-rate magnitude is estimated as:
\begin{equation}
    |S| \sim \frac{U_0}{L^{1/3} \ell^{2/3}}, \quad L = N_{x, LES}, \quad \ell = 1 \text{ (the smallest resolved lattice scale, Sec.~\ref{sec:hydrolimit})},
\end{equation}
yielding $|S| \approx 0.00926$. The turbulent relaxation-time correction is then
\begin{equation}
    \tau_t = \frac{C \Delta^2}{c_s^2} |S| = \frac{0.13 \cdot 1^2}{1/3} \cdot 0.00926 = 0.0036114,
\end{equation}
so that
\begin{equation}
    \frac{\tau_t}{\tau_{0,LES} - 0.5} \approx 5.6, \qquad
    \tau_{e} \approx 0.5 + (\tau_{0,LES} - 0.5) + \tau_t = 0.5042564,
\end{equation}
with
$\nu_{0, DNS} = 0.00172$
\begin{equation}
    \nu_e = c_s^2 (\tau_{e} - 0.5) \approx 0.0014188, \qquad
    Re_e \approx \frac{U_0 N_{x, LES}}{\nu_e} \approx 3030.
\end{equation}
This K41-based estimate provides a conservative upper bound;
the measured offsets are indeed smaller, as verified by the distribution
shown in Fig.~\ref{fig:tau_histograms}.

\subsection{Effective relaxation-time analysis}
\label{sec:effective_relaxation_time_analysis}

\begin{figure}[b]
    \centerline{\includegraphics[width=9.4cm]{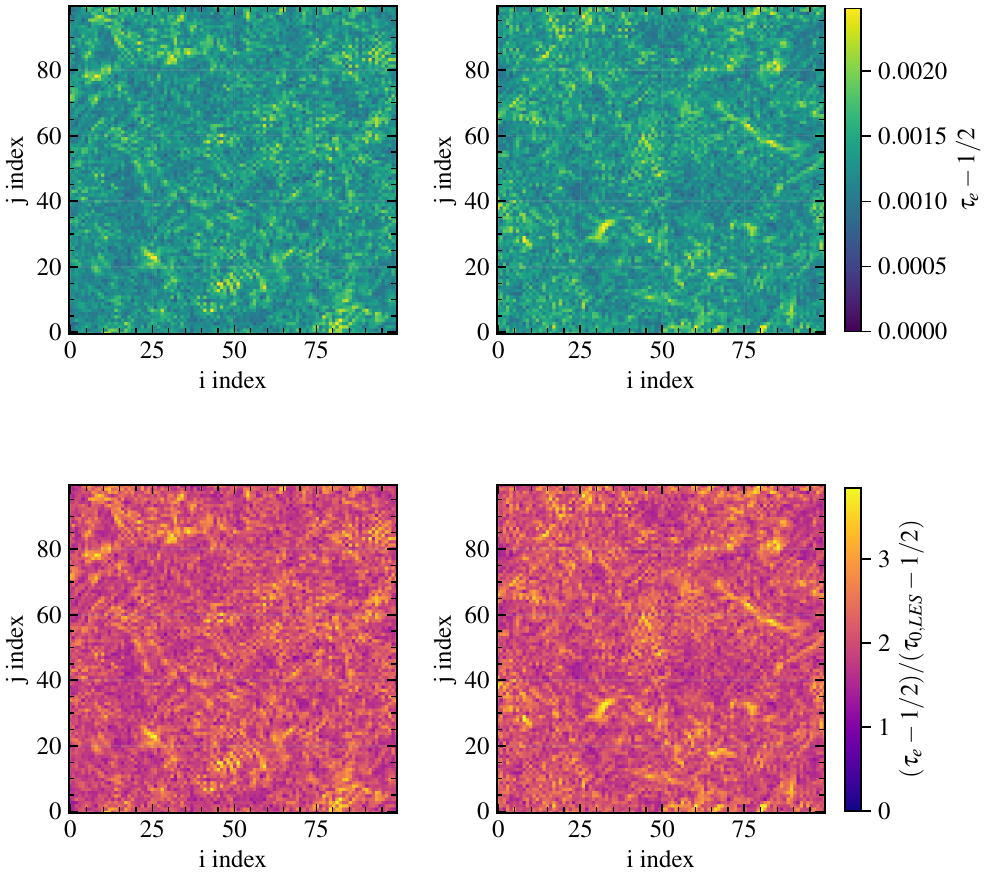}}
    \vspace*{8pt}
    \caption{Operating-regime maps at two non-dimensional times $t/t_0=200,\,600$ on slice $k=50$. Top: absolute offset $\tau_e-\tfrac{1}{2}$ (common color limits $0$-$2.5\times10^{-3}$; rare peaks up to $\sim3.7\times10^{-3}$ are clipped). Bottom: normalized offset $(\tau_e-\tfrac{1}{2})/(\tau_{0,\text{LES}}-\tfrac{1}{2})$ (common limits $0$-$3.85$; rare peaks up to $\sim5.7$ are clipped). Typical slice statistics across the selected snapshots: mean $\approx 1.34\times10^{-3}$ (normalized $\approx 2.08$), 95th percentile $\lesssim 1.82\times10^{-3}$ (normalized $\lesssim 2.83$). Values $\geq 1$ confirm turbulent enhancement over the molecular baseline.}
    \label{fig:tau_maps}
\end{figure}

To directly diagnose the operating regime, we visualize the effective relaxation-time offset $\tau_e-\tfrac{1}{2}=3(\nu_0+\nu_t)$. Fig.~\ref{fig:tau_maps} shows spatial maps of $\tau_e-\tfrac{1}{2}$ (top row) and the LES-normalized field $(\tau_e-\tfrac{1}{2})/(\tau_{0,\mathrm{LES}}-\tfrac{1}{2})$ (bottom row) at two representative times on a fixed slice. Color limits are fixed across columns (absolute: $0$-$2.5\times10^{-3}$; normalized: $0$-$3.85$ with rare peaks up to $\sim5.7$ clipped) to enable direct comparison.

The fields are spatially intermittent but small: typical slice statistics are mean $\approx 1.30\times10^{-3}$, median $\approx 1.31\times10^{-3}$, 95th percentile $\approx 1.81\times10^{-3}$, 99th $\approx 2.09\times10^{-3}$, with rare slice maxima $\approx 2.65\times10^{-3}$ (global maxima across the analyzed volume/time up to $\approx 3.68\times10^{-3}$). In normalized units the slice mean is $\approx 2.01$, 95th $\approx 2.81$, 99th $\approx 3.25$ (global max $\approx 5.70$).
As derived in Sec.~\ref{sec:hydrolimit}, the hydrodynamic limit for the present LES requires $\tau-\tfrac{1}{2}\ll\mathcal{O}(1)$, so the measured offsets are approximately three orders of magnitude smaller and therefore firmly hydrodynamic.

Since the LES molecular baseline is $3\nu_0=6.45\times10^{-4}$, the normalized ratio shows turbulent contributions are $\sim 2$-$3\times$ the molecular baseline, confirming $\tau_e = \tau_0 + \tau_t$ and evidencing operation in the hydrodynamic regime $\tau=\tfrac{1}{2}+\theta$.

\begin{figure}[b]
    \centerline{\includegraphics[width=9.4cm]{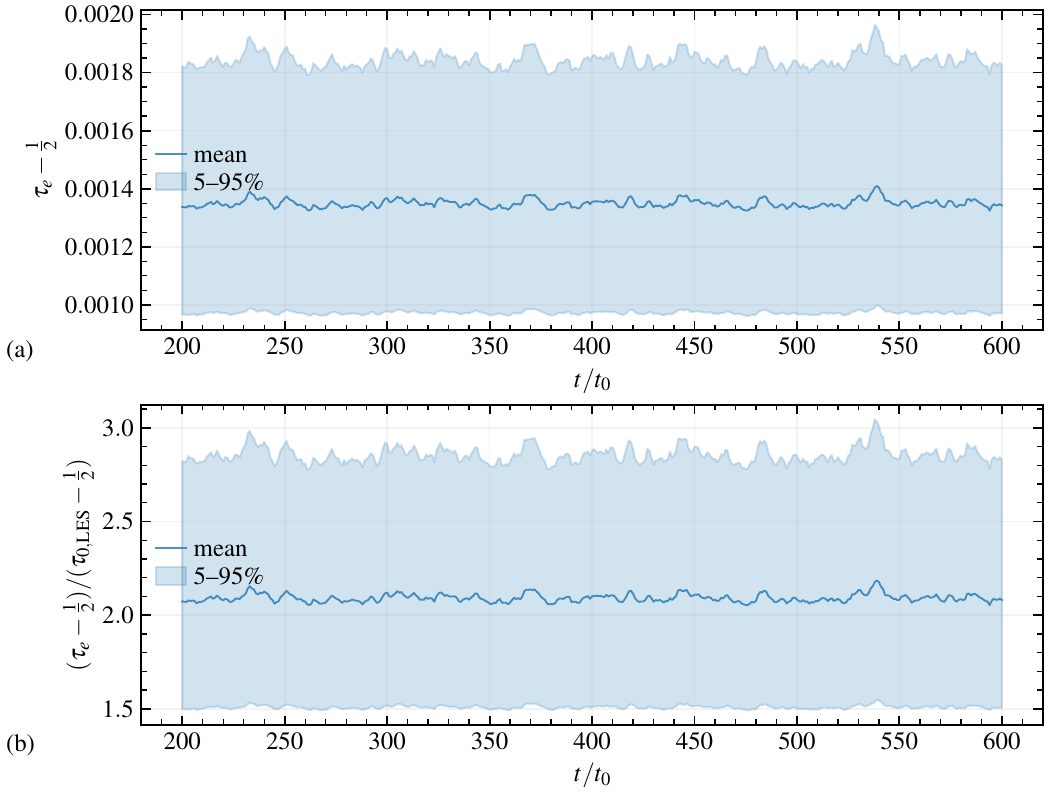}}
    \vspace*{8pt}
    \caption{Temporal statistics of the effective relaxation-time offset. Top: domain mean (solid) and 5-95\% envelope (shaded) of \(\tau_e-\tfrac{1}{2}\). Bottom: normalized counterpart \((\tau_e-\tfrac{1}{2})/(\tau_{0,\text{LES}}-\tfrac{1}{2})\). Means \(\approx 1.34\times10^{-3}\) and \(\approx 2.08\) persist in time; the 95th percentile remains \(\lesssim 1.82\times10^{-3}\) (normalized \(\lesssim 2.83\)), confirming a stable hydrodynamic regime with no negative excursions.}
    \label{fig:tau_timeseries}
\end{figure}

Figure~\ref{fig:tau_timeseries} reports the domain-mean \(\langle\tau_e-\tfrac{1}{2}\rangle\) and its 5-95\% envelope (top) together with the normalized counterpart (bottom). The mean remains nearly constant at \(\approx 1.34\times10^{-3}\) (normalized \(\approx 2.08\)), and the 95th percentile stays below \(\approx 1.82\times10^{-3}\) (normalized \(\approx 2.83\)) over the analyzed window. Across all snapshots we found no negative values of \(\tau_e-\tfrac{1}{2}\); the 5th percentile remained \(>\!1.0\times10^{-3}\) (normalized \(>\!1.55\)), consistent with expectations for homogeneous turbulence.

From the regime diagnostics (Fig.~\ref{fig:tau_maps}-\ref{fig:tau_timeseries}), the domain-mean offset is $\langle\tau_e-\tfrac{1}{2}\rangle \approx 1.34\times10^{-3}$ with a 95th percentile $\lesssim 1.82\times10^{-3}$ over $t/t_0\in[200,600]$, i.e.\ $\tau_e \approx 0.5013$ with no negative excursions. Offsets are only a few multiples of the molecular baseline $3\nu_0=6.45\times10^{-4}$ and $\ll 1$, so collisions remain in the hydrodynamic regime $\tau=\tfrac{1}{2}+\theta$ with $\theta=\mathcal{O}(10^{-3})$, far below the formal hydrodynamic limit (Sec.~\ref{sec:hydrolimit}).
This ensures:
\begin{itemlist}
    \item collisions remain strictly local (no remote data dependencies);
    \item the scheme stays well within LBM stability ($\tau>0.5$) \cite{kruger2017lattice};
    \item post-collision distributions remain bounded and physical.
\end{itemlist}

\subsection{Spectral Comparison}
\label{sec:spectral_analysis}

\begin{figure}[b]
    \centerline{\includegraphics[width=9.4cm]{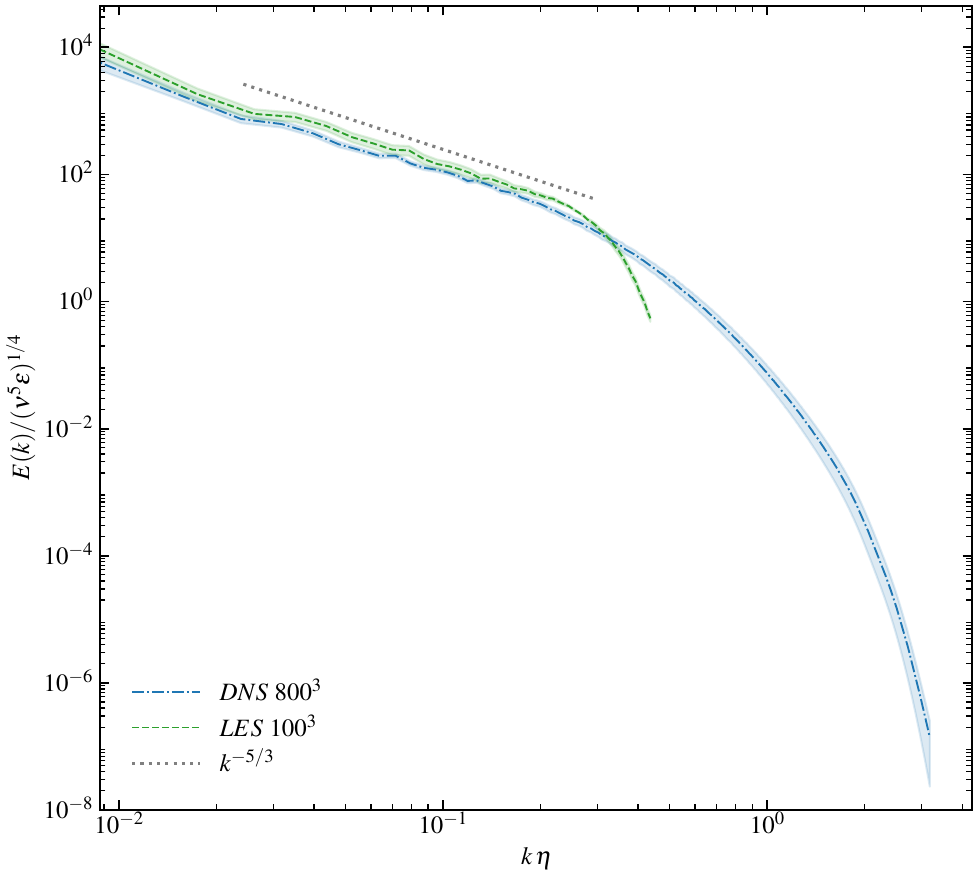}}
    \vspace*{8pt}
    \caption{Comparison of $E(k)$ from DNS ($800^3$, $Re_\lambda \approx 243$, $\eta^+ \approx 1.001$) and LES ($100^3$), plotted against $k\eta$ and normalized by $(\nu \epsilon)^{1/4}$. The dashed line indicates the reference $k^{-5/3}$ scaling.}
    \label{fig:Average_spectra_comparison_800}
    \alttext{Comparison of energy spectra from DNS and LES simulations showing k^{-5/3} scaling.}
\end{figure}

The energy spectra in Fig.~\ref{fig:Average_spectra_comparison_800} validate the physical consistency of the LES-MRT formulation. While the DNS spectrum ($800^3$) serves as reference, the LES ($100^3$) exhibits the canonical $k^{-5/3}$ scaling throughout the resolved inertial range (approximately $k\eta \lesssim 3 \times 10^{-1}$), confirming correct large-scale transfer. The spectral roll-off observed at $k\eta \approx 4 \times 10^{-1}$ reflects filtering at the LES grid scale and subgrid dissipation, which truncate the small scales. Crucially, there is no non-physical energy pile-up near the cut-off, indicating adequate subgrid dissipation. This is consistent with the small measured offsets ($\theta=\mathcal{O}(10^{-3})$) that confirm operation in the hydrodynamic regime, where spatial locality is preserved and the Smagorinsky model functions correctly.

\subsection{Structure Function and Flatness Comparison}
\label{sec:high_order_stats}

\begin{figure}[b]
    \centerline{\includegraphics[width=9.4cm]{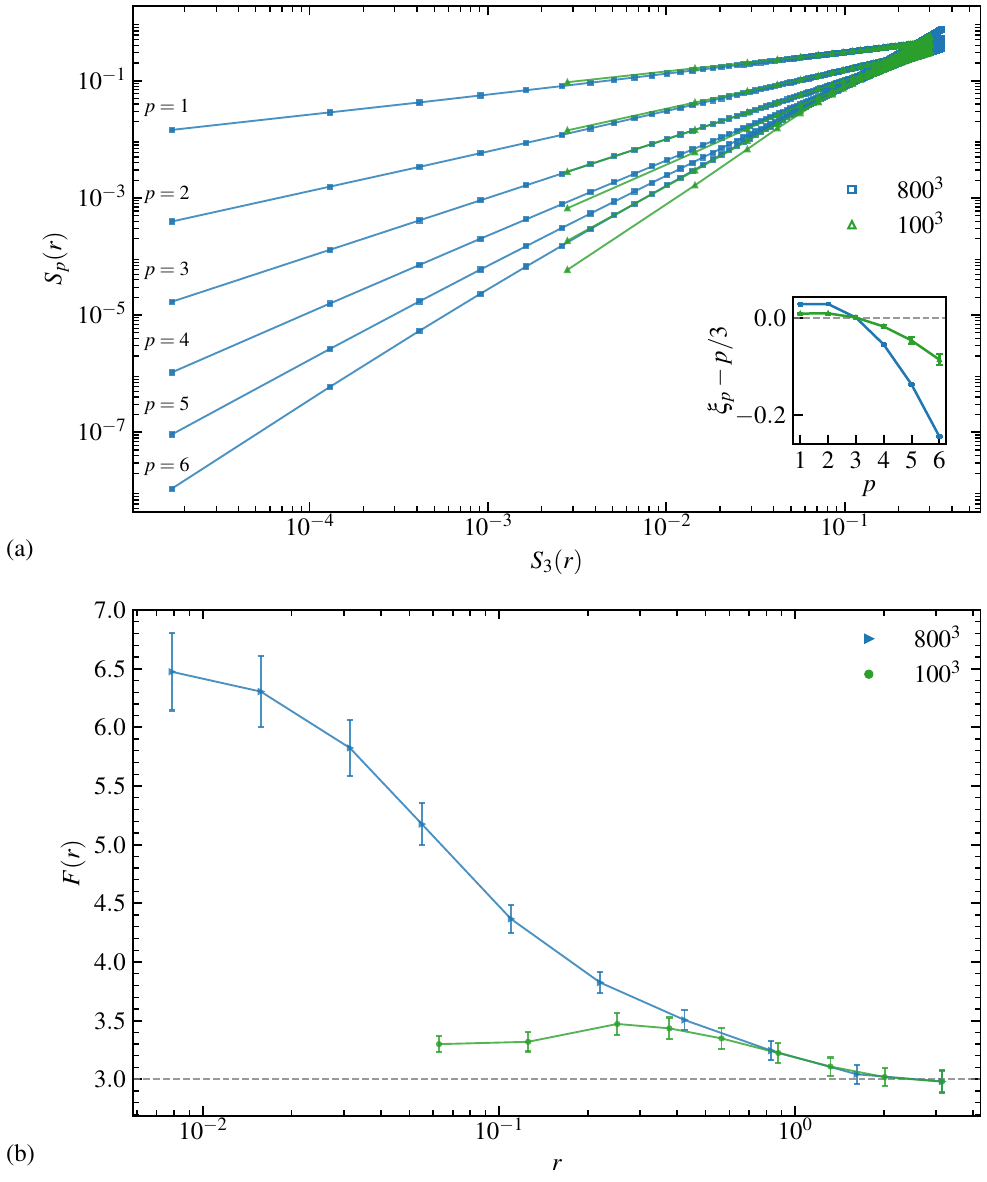}}
    \vspace*{8pt}
    \caption{Comparison of higher-order statistics between LES-MRT ($100^3$) and reference DNS ($800^3$).
        (a) ESS scaling of structure functions $S_p(r)$ shows good collapse for lower orders ($p \leq 3$), while higher-order exponents deviate in LES due to grid coarsening and subgrid modeling near the filter scale.
        (b) Flatness $F_4(r)$ displays reduced small-scale intermittency in LES, consistent with Smagorinsky damping near the filter scale. Both trends are consistent with standard LES behavior and confirm that spatial locality is preserved.}
    \label{fig:les_dns_flatness_structure}
\end{figure}

Further validation through ESS and $F_4(r)$ (Fig.~\ref{fig:les_dns_flatness_structure}) confirms the physical consistency of the LES-MRT approach. The ESS analysis (Fig.~\ref{fig:les_dns_flatness_structure} (a)) demonstrates excellent collapse with DNS for structure functions up to third order ($p \leq 3$) across resolved inertial scales. However, higher-order moments ($p \geq 4$) exhibit scaling exponents closer to non-intermittent K41 scaling (see Inset, smaller deviation $\xi_p - p/3$), a direct consequence of grid coarsening ($\delta x_{\text{LES}} = 8\delta x_{\text{DNS}}$) filtering subgrid vortices.

Corroborating this, the flatness $F_4(r)$ (Fig.~\ref{fig:les_dns_flatness_structure}(b)) exhibits the characteristic behaviour expected in LES: while agreement with DNS is retained at large scales, a noticeable reduction in peak flatness appears as $r$ approaches the LES grid scale $\delta x_{\text{LES}}$, reflecting the attenuation of extreme intermittent events \cite{buzzicotti2021inertial}. These deviations are confined to unresolved scales and show no evidence of locality breakdown: $F_4(r)$ remains bounded, the relaxation offset is small ($\tau_e-\tfrac12=\mathcal{O}(10^{-3})$), and the trends mirror standard LES phenomenology. Collectively, these results verify that the LBM-MRT scheme preserves spatial locality while reproducing essential turbulence statistics of conventional LES methodologies.

These statistical comparisons reinforce the spectral findings and confirm that, while some under-resolution at small scales is inherent to any LES approach, the MRT-based LBM formulation reproduces the key characteristics of homogeneous incompressible turbulent flow. The effective relaxation time remains very close to $0.5$: the domain-mean offset is $\langle\tau_e-\frac{1}{2}\rangle \approx 1.34\times10^{-3}$ (95th percentile $\lesssim 1.82\times10^{-3}$), i.e.\ $\tau_e \approx 0.5013$, with no negative excursions. The normalized ratio shows turbulent contributions are $\sim 2$–$3\times$ the molecular baseline $3\nu_0=6.45\times10^{-4}$ (normalized mean $\approx 2.08$), ensuring collision-streaming remains strictly local despite modeled subgrid dissipation. The observed agreement in energy spectra, structure-function scaling, and flatness validates that the scheme behaves as a standard large-eddy simulation while preserving the spatial locality and stability intrinsic to LBM.

\clearpage

\appendix

\section{LBM-MRT Governing Equations and Forcing}
\label{app:lbe_mrt_forcing}

For completeness, we collect here the governing equations and forcing used in the MRT-LBM formulation. The mesoscopic lattice Boltzmann equation (LBE) is defined as:
\begin{equation}
    \frac{\partial f(x, \xi, t)}{\partial t} + \xi \frac{\partial f(x, \xi, t)}{\partial x} + F \frac{\partial f(x, \xi, t)}{\partial \xi} = \Omega(f(x, \xi, t)),
    \label{eq:continues_eq}
\end{equation}

\noindent where $f(x, \xi, t)$ represents the particle distribution function, $F$ is the external force, and $\Omega$ the collision operator.

\begin{figure}[b]
    \centerline{

        %---------- (a) left panel: tabular-as-figure ----------
        \begin{subfigure}{0.49\linewidth}
            \centering
            \footnotesize % keeps it readable after reduction
            \setlength{\tabcolsep}{4pt} % a bit tighter columns
            \begin{tabular}{ccc}
                \toprule
                \textbf{Index} & $c_{\alpha}$ & $w_{\alpha}$ \\
                \colrule
                1              & (1, 0,   0)  & 1/18         \\
                2              & (-1, 0,  0)  & 1/18         \\
                3              & (0, 1,   0)  & 1/18         \\
                4              & (0, -1,  0)  & 1/18         \\
                5              & (0, 0,   1)  & 1/18         \\
                6              & (0, 0, - 1)  & 1/18         \\
                7              & (1, 1,   0)  & 1/36         \\
                8              & (-1, -1, 0)  & 1/36         \\
                9              & (1, -1,  0)  & 1/36         \\
                10             & (-1, 1,  0)  & 1/36         \\
                11             & (0, 1,   1)  & 1/36         \\
                12             & (0, -1, -1)  & 1/36         \\
                13             & (0, 1,  -1)  & 1/36         \\
                14             & (0, -1,  1)  & 1/36         \\
                15             & (1, 0,   1)  & 1/36         \\
                16             & (-1, 0, -1)  & 1/36         \\
                17             & (1, 0,  -1)  & 1/36         \\
                18             & (-1, 0,  1)  & 1/36         \\
                19             & (0, 0,   0)  & 1/3          \\
                \botrule
            \end{tabular}
            \caption{}
            \label{fig:d3q19_table}
        \end{subfigure}
        \hfill

        %---------- (b) right panel: image ----------
        \begin{subfigure}{0.49\linewidth}
            \centering
            \includegraphics[width=9.4cm]{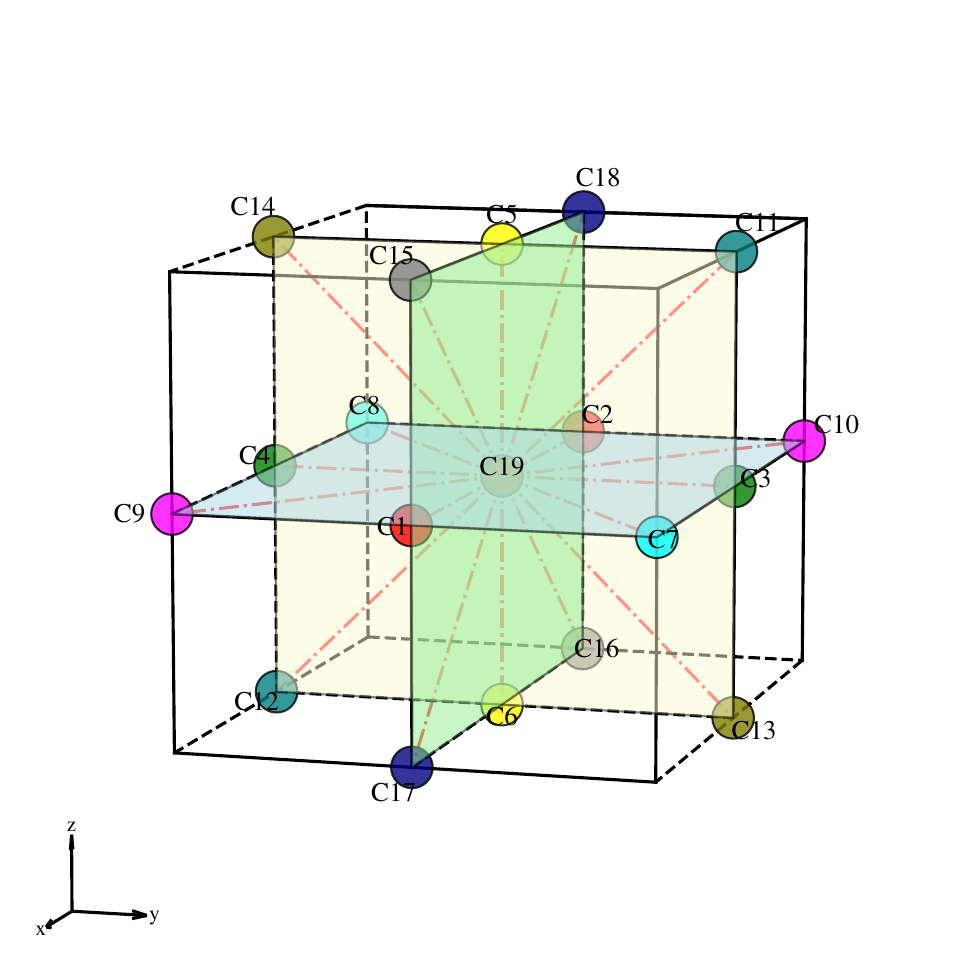}
            \vspace*{8pt}
            \caption{}
            \label{fig:d3q19_stencil}
        \end{subfigure}
    }
    \caption{(a) D3Q19 velocity set $c_\alpha$ and weights $w_\alpha$; (b) D3Q19 velocity stencil.}
    \label{fig:d3q19-combined}
\end{figure}

To solve it numerically, we discretize in time and velocity space using the D3Q19 stencil (see Fig.~\ref{fig:d3q19_stencil}, Fig.~\ref{fig:d3q19_table}). We employ the MRT formulation \cite{d2002multiple}, which provides enhanced stability over BGK. Incorporating Guo's forcing scheme \cite{guo2002discrete}, the discrete evolution equation is:
\begin{equation}
    f(x + c_\alpha \delta t, t + \delta t) - f(x, t) =
    - \left[ M^{-1} \Lambda M (f - f^{eq}) \right]
    + \delta t \left[ M^{-1} \left( I - \frac{1}{2} \Lambda \right) M \right] \Phi,
    \label{eq:mrt_dns_guo}
\end{equation}

\noindent where $M$ is the moment-transformation matrix and $\Lambda$ the diagonal matrix of relaxation rates.

The equilibrium distribution function \cite{he1997lattice} is:
\begin{equation}
    f^{eq}_\alpha = w_\alpha \rho \left[
        1 + \frac{c_\alpha \cdot u}{c_s^2}
        + \frac{(c_\alpha \cdot u)^2}{2 c_s^4}
        - \frac{u \cdot u}{2c_s^2} \right],
    \label{eq:feq}
\end{equation}

The Guo forcing term is
\begin{equation}
    \Phi_\alpha = w_\alpha \left[ \frac{c_\alpha - u}{c_s^2} + \frac{(c_\alpha \cdot u) c_\alpha}{c_s^4} \right] \cdot F^{ext},
    \label{eq:guo_force}
\end{equation}

The corrected velocity is
\begin{equation}
    u = \frac{1}{\rho} \left( \sum_\alpha f_\alpha c_\alpha + \frac{\delta t}{2} F^{ext} \right),
    \label{eq:guo_velocity}
\end{equation}

and macroscopic quantities are
\begin{equation}
    \rho = \sum_\alpha f_\alpha, \quad \rho u = \sum_\alpha f_\alpha c_\alpha + \frac{\delta t}{2}F^{ext}.
    \label{eq:macroscopic_eq}
\end{equation}

The lattice speed is $c=\delta x/\delta t$, with $\delta x=1$, $\delta t=1$ in lattice units, and the sound speed is $c_s=c/\sqrt{3}$ for D3Q19 \cite{mohamad2011lattice}.

\section{MRT Model Parameters}
\label{app:mrt_parameters}

The transformation matrix $M$ for the D3Q19 lattice is constructed using the orthogonal moment basis \cite{d2002multiple}

\begin{equation}
    M = \scalebox{0.7}{$\left[
                \begin{array}{*{19}{r}}
                    1   & 1   & 1   & 1   & 1   & 1   & 1  & 1  & 1  & 1  & 1  & 1  & 1  & 1  & 1  & 1  & 1  & 1  & 1   \\
                    -11 & -11 & -11 & -11 & -11 & -11 & 8  & 8  & 8  & 8  & 8  & 8  & 8  & 8  & 8  & 8  & 8  & 8  & -30 \\
                    -4  & -4  & -4  & -4  & -4  & -4  & 1  & 1  & 1  & 1  & 1  & 1  & 1  & 1  & 1  & 1  & 1  & 1  & 12  \\
                    1   & -1  & 0   & 0   & 0   & 0   & 1  & -1 & 1  & -1 & 0  & 0  & 0  & 0  & 1  & -1 & 1  & -1 & 0   \\
                    -4  & 4   & 0   & 0   & 0   & 0   & 1  & -1 & 1  & -1 & 0  & 0  & 0  & 0  & 1  & -1 & 1  & -1 & 0   \\
                    0   & 0   & 1   & -1  & 0   & 0   & 1  & -1 & -1 & 1  & 1  & -1 & 1  & -1 & 0  & 0  & 0  & 0  & 0   \\
                    0   & 0   & -4  & 4   & 0   & 0   & 1  & -1 & -1 & 1  & 1  & -1 & 1  & -1 & 0  & 0  & 0  & 0  & 0   \\
                    0   & 0   & 0   & 0   & 1   & -1  & 0  & 0  & 0  & 0  & 1  & -1 & -1 & 1  & 1  & -1 & -1 & 1  & 0   \\
                    0   & 0   & 0   & 0   & -4  & 4   & 0  & 0  & 0  & 0  & 1  & -1 & -1 & 1  & 1  & -1 & -1 & 1  & 0   \\
                    2   & 2   & -1  & -1  & -1  & -1  & 1  & 1  & 1  & 1  & -2 & -2 & -2 & -2 & 1  & 1  & 1  & 1  & 0   \\
                    -4  & -4  & 2   & 2   & 2   & 2   & 1  & 1  & 1  & 1  & -2 & -2 & -2 & -2 & 1  & 1  & 1  & 1  & 0   \\
                    0   & 0   & 1   & 1   & -1  & -1  & 1  & 1  & 1  & 1  & 0  & 0  & 0  & 0  & -1 & -1 & -1 & -1 & 0   \\
                    0   & 0   & -2  & -2  & 2   & 2   & 1  & 1  & 1  & 1  & 0  & 0  & 0  & 0  & -1 & -1 & -1 & -1 & 0   \\
                    0   & 0   & 0   & 0   & 0   & 0   & 1  & 1  & -1 & -1 & 0  & 0  & 0  & 0  & 0  & 0  & 0  & 0  & 0   \\
                    0   & 0   & 0   & 0   & 0   & 0   & 0  & 0  & 0  & 0  & 1  & 1  & -1 & -1 & 0  & 0  & 0  & 0  & 0   \\
                    0   & 0   & 0   & 0   & 0   & 0   & 0  & 0  & 0  & 0  & 0  & 0  & 0  & 0  & 1  & 1  & -1 & -1 & 0   \\
                    0   & 0   & 0   & 0   & 0   & 0   & 1  & -1 & 1  & -1 & 0  & 0  & 0  & 0  & -1 & 1  & -1 & 1  & 0   \\
                    0   & 0   & 0   & 0   & 0   & 0   & -1 & 1  & 1  & -1 & 1  & -1 & 1  & -1 & 0  & 0  & 0  & 0  & 0   \\
                    0   & 0   & 0   & 0   & 0   & 0   & 0  & 0  & 0  & 0  & -1 & 1  & 1  & -1 & 1  & -1 & -1 & 1  & 0
                \end{array}
                \right]$}
    \label{eq:moment-matrix}
\end{equation}

%%%%%%%%%%%%% Relaxation matrix%%%%%%%%
The diagonal relaxation matrix is
\begin{equation}
    S \equiv \mathrm{diag}(1.0, 1.19, 1.4, 1.0, 1.2, 1.0, 1.2, 1.0, 1.2, s_\nu, 1.4, s_\nu, 1.4, s_\nu, s_\nu, s_\nu, 1.98, 1.98, 1.98)
    \label{eq:relaxation-matrix}
\end{equation}

%%%%%%%%%%%%%%%%%% Moment and equilibrium moments
The equilibrium and forcing moments for D3Q19 LBM-MRT model are given by:
\begin{equation}
    \begin{aligned}
        m^{(eq)} & =
        \left[\begin{array}{c}
                      \delta\rho = \rho - \rho_0                    \\
                      -11\delta\rho + 19\rho(u_x^2 + u_y^2 + u_z^2) \\
                      -\frac{475}{63}\rho(u_x^2 + u_y^2 + u_z^2)    \\
                      \rho u_x                                      \\
                      -\frac{2\rho u_x}{3}                          \\
                      \rho u_y                                      \\
                      -\frac{2\rho u_y}{3}                          \\
                      \rho u_z                                      \\
                      -\frac{2\rho u_z}{3}                          \\
                      \rho(2u_x^2 - u_y^2 - u_z^2)                  \\
                      0                                             \\
                      \rho(u_y^2 - u_z^2)                           \\
                      0                                             \\
                      \rho u_x u_y                                  \\
                      \rho u_y u_z                                  \\
                      \rho u_x u_z                                  \\
                      0                                             \\
                      0                                             \\
                      0
                  \end{array}\right],
        \quad
        F_m =
        \left[\begin{array}{c}
                      0                                \\
                      38(u_x F_x + u_y F_y + u_z F_z)  \\
                      -11(u_x F_x + u_y F_y + u_z F_z) \\
                      F_x                              \\
                      -\frac{2F_x}{3}                  \\
                      F_y                              \\
                      -\frac{2F_y}{3}                  \\
                      F_z                              \\
                      -\frac{2F_z}{3}                  \\
                      2(2u_x F_x - u_y F_y - u_z F_z)  \\
                      -(2u_x F_x - u_y F_y - u_z F_z)  \\
                      2(u_y F_y - u_z F_z)             \\
                      -(u_y F_y - u_z F_z)             \\
                      u_x F_y + u_y F_x                \\
                      u_y F_z + u_z F_y                \\
                      u_x F_z + u_z F_x                \\
                      0                                \\
                      0                                \\
                      0
                  \end{array}\right]
    \end{aligned}
    \label{eq:meq_fm}
\end{equation}

\subsection{Strain Rate Tensor in LBM-MRT}
\label{app:strain_rate_tensor}

The components of the filtered strain-rate tensor used in the LES-MRT formulation are computed from the non-equilibrium moments as follows \cite{yu2005dns}. Here, $\rho_0$ denotes the reference density (set to unity in lattice units for incompressible flow):
{\small
\begin{align}
    S_{xx} & = -\frac{s_1 m_1^{(neq)} + 19s_9 m_9^{(neq)}}{38\rho_0\delta_t}, \quad
    S_{yy} = -\frac{2s_1 m_1^{(neq)} - 19s_9(m_9^{(neq)} - 3m_{11}^{(neq)})}{76\rho_0\delta_t}, \nonumber \\
    S_{zz} & = -\frac{2s_1 m_1^{(neq)} - 19s_9(m_9^{(neq)} + 3m_{11}^{(neq)})}{76\rho_0\delta_t}, \quad
    S_{xy} = -\frac{3s_9}{2\rho_0\delta_t} m_{13}^{(neq)}, \nonumber                                      \\
    S_{xz} & = -\frac{3s_9}{2\rho_0\delta_t} m_{15}^{(neq)}, \quad
    S_{yz} = -\frac{3s_9}{2\rho_0\delta_t} m_{14}^{(neq)}
    \label{eq:yu_strain}
\end{align}
}

\subsection{Distribution of effective relaxation time offsets}
\label{app:tau_histograms}

To confirm that the MRT-LES scheme operates within the theoretically predicted regime,
we examine the distribution of effective relaxation time offsets
$\tau_e - \tfrac{1}{2}$ across the lattice.
From the scaling analysis in Sec.~\ref{sec:results}, the expected mean offset is
\begin{equation}
    \tau_e - \frac{1}{2} \approx 0.00426,
\end{equation}

which provides a conservative upper bound for the scheme's operating regime.

\begin{figure}[b]
    \centerline{\includegraphics[width=9.4cm]{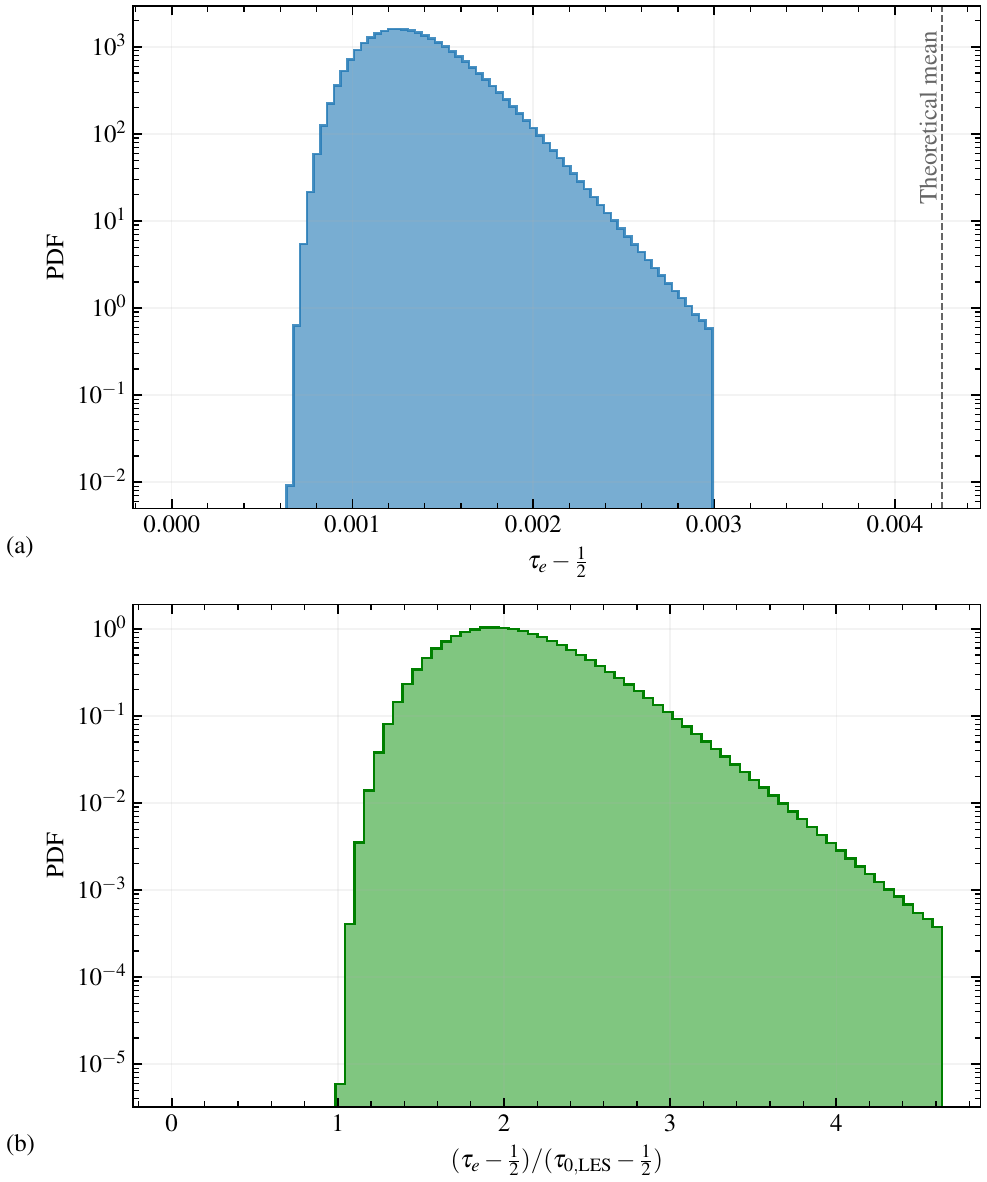}}
    \vspace*{8pt}
    \caption{Distribution of effective relaxation time offsets in MRT-LES simulations.
        (a) Top: absolute histogram of $\tau_e - \frac{1}{2}$ with theoretical mean $\approx 0.00426$ (dashed line),
        showing measured values remain below the conservative theoretical bound.
        (b) Bottom: LES-normalized histogram $(\tau_e - \tfrac{1}{2})/(\tau_{0,\text{LES}} - \tfrac{1}{2})$;
        values are $\ge 1$ by construction with a bulk around $\sim 2$-$3$, consistent with the time-series.}
    \label{fig:tau_histograms}
\end{figure}

Fig.~\ref{fig:tau_histograms} shows histograms of the measured offsets in absolute and LES-normalized form.
The absolute distribution (left) peaks in the range $[10^{-3},\,3\times10^{-3}]$ and lies entirely below the
theoretical upper bound, confirming stability. The LES-normalized distribution (right),
$(\tau_e - \tfrac{1}{2})/(\tau_{0,\text{LES}} - \tfrac{1}{2})$, spans values $\gtrsim 1$ with a bulk around $\sim 2$-$3$,
consistent with $\tau_e = \tau_0 + \tau_t$ and the time-series statistics.

\section{Additional Validation Results}
\subsection{Isotropy}
\label{app:spectral_isotropy}

\begin{figure}[b]
    \centerline{\includegraphics[width=9.4cm]{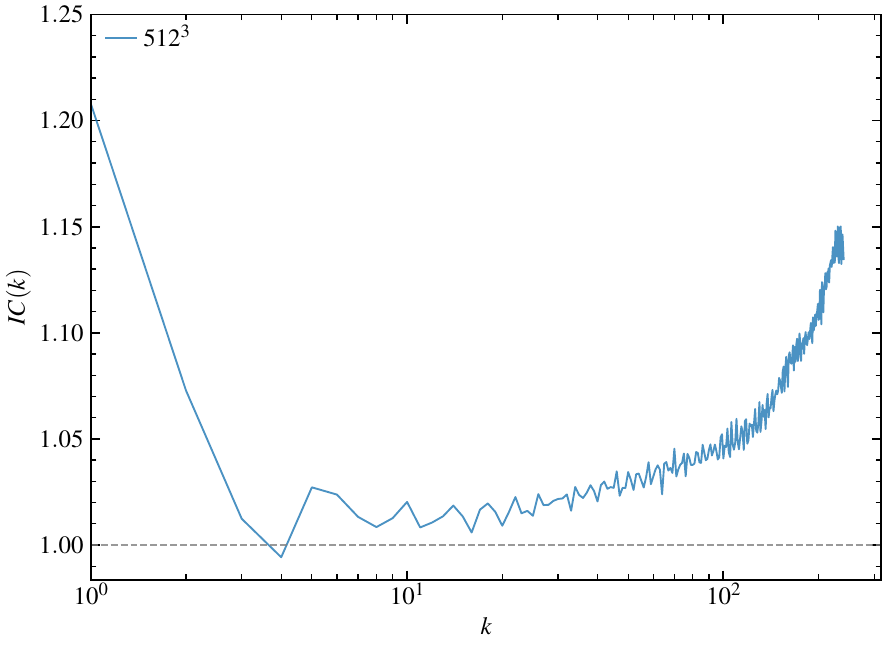}}
    \vspace*{8pt}
    \caption{Isotropy coefficient for $512^3$ ($Re_\lambda \approx 180$, $\eta^+ \approx 1.0018$) simulation, time-averaged over $t/t_0 = 100$-$1000$. The theoretical isotropic value ($IC = 1$, dashed) is closely matched, confirming homogeneous isotropy.}
    \label{fig:spectral_IC}
\end{figure}

The isotropy coefficient $IC(k)$ quantifies directional uniformity in turbulence \cite{batchelor1953theory}. Following \cite{singh2024comparison}, we define:
\begin{equation}
    IC(k) = \frac{2E_{22}(k) - k \frac{\partial E_{11}}{\partial k}}{2E_{11}(k)}
    \label{eq:isotropy_coefficient}
\end{equation}
where $E_{11}$ and $E_{22}$ denote longitudinal and transverse energy spectra. Fig.~\ref{fig:spectral_IC} shows $IC(k) \approx 1$ across resolved wavenumbers, confirming statistical isotropy consistent with Sec.~\ref{subsec:isotropy}.

\subsection{DNS-LES Spectra Comparison}
\label{app:dns_les_spectra}

\begin{figure}[b]
    \centerline{\includegraphics[width=9.4cm]{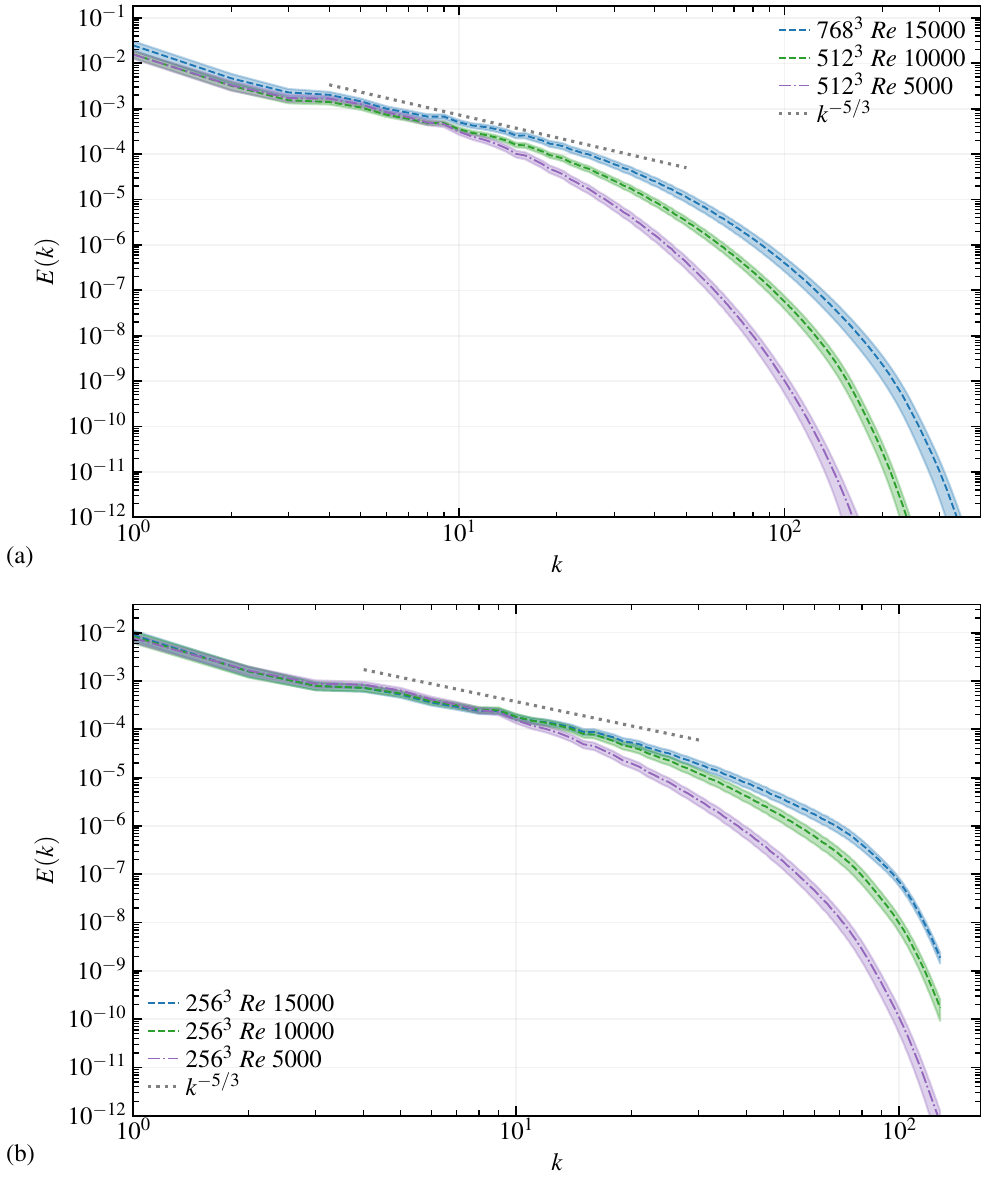}}
    \vspace*{8pt}
    \caption{$E(k)$ from DNS (a) and LES (b) simulations across different $Re$ and resolutions. LES results exhibit consistent inertial-range scaling.
    }
    \label{fig:combined_dns_les_spectra}
\end{figure}

\clearpage
\section*{Acknowledgments}
The authors wish to acknowledge the support of the National Center for HPC, Big Data and Quantum Computing, Project CN\_00000013---CUP E83C22003230001, Mission 4 Component 2 Investment 1.4, funded by the European Union---NextGenerationEU; the support of Project PRIN 2022F422R2---CUP E53D23003210006, financed by the European Union---Next Generation EU; the support of Project PRIN PNRR P202298P25---CUP E53D23016990001, financed by the European Union---Next Generation EU; and the support of Project ECS 0000024 Rome Technopole---CUP B83C22002820006, NRP Mission 4 Component 2 Investment 1.5, funded by the European Union---NextGenerationEU.

The authors gratefully acknowledge Prof.~Luca Biferale for valuable discussions and insightful feedback that significantly improved this work.

The authors used Grammarly's AI writing assistant to refine grammar and improve readability of the manuscript, and Cursor AI to assist in identifying bugs in the code. These tools were used solely for language polishing and technical debugging. All content was subsequently reviewed, verified, and revised by the authors, who take full responsibility for the final version of this work.

\bibliographystyle{ws-ijmpc}
\bibliography{references}
\end{document}